\documentclass[10pt]{article}
\usepackage{flafter,amsmath,amssymb,latexsym,psfrag,graphicx,color}
\usepackage[margin=2.5cm]{geometry}
\usepackage[numbers,sort&compress]{natbib}

\newtheorem{theorem}{Theorem}[section]

\newtheorem{lemma}[theorem]{Lemma}
\newtheorem{remark}[theorem]{Remark}

\makeatletter

\@addtoreset{figure}{section}
\makeatother

\begin{document}
\setlength\arraycolsep{2pt}
\date{\today}

\title{Reconstruction of an unknown cavity with Robin boundary condition inside a heat conductor}
\author{Gen Nakamura$^1$, Haibing Wang$^{2}$\footnote{Corresponding author, E-mail: hbwang@seu.edu.cn}\\
\\$^1$Department of Mathematics, Inha University, Incheon 402-751, Korea \quad\quad\quad\quad\quad
\\$^2$Department of Mathematics, Southeast University, Nanjing 210096, P.R. China
}

\maketitle
\begin{abstract}
Active thermography is a non-destructive testing technique to detect the internal structure of a heat conductor, which is widely applied in industrial engineering. In this paper, we consider the problem of identifying an unknown cavity with Robin boundary condition inside a heat conductor from boundary measurements. To set up the inverse problem mathematically, we first state the corresponding forward problem and show its well-posedness in an anisotropic Sobolev space by the integral equation method. Then, taking the Neumann-to-Dirichlet map as mathematically idealized measured data for the active thermography, we present a linear sampling method for reconstructing the unknown Robin-type cavity and give its mathematical justification by using the layer potential argument. In addition, we analyze the indicator function used in this method and show its pointwise asymptotic behavior by investigating the reflected solution of the fundamental solution. From our asymptotic analysis, we can establish a pointwise reconstruction scheme for the boundary of the cavity, and can also know the distance to the unknown cavity as we probe it from its inside.

\bigskip



\end{abstract}

\section{Introduction}
\setcounter{equation}{0}

Active thermography is a widely used non-destructive testing technique in industrial engineering, which aims to detect the internal information of a heat conductor \cite{C-M1992, Iba2009, P-L-A1992, R-N-M2003}. The principle of active thermography is as follows. If there is an anomaly inside the conductor, it affects the propagation of the heat flow inside the conductor and as a result it also directly affects the temporal behavior of the surface temperature distribution. By measuring the surface temperature distribution and by solving some inverse problem, the information on the anomaly can be calculated. The information we want to know are the size, location and shape of anomalies and their physical properties such as heat conductivities. The measurement of active thermography is a non-contact, very fast and large area measurement which is conducted by injecting a heat flux to the conductor by a flash lamp or heater and measuring the corresponding distribution of temperature on the surface of the conductor by an infrared light camera.

In this paper, we want to recover an unknown cavity with Robin boundary condition inside a heat conductor via active thermography. To begin with, we give the mathematical formulation as follows. Let $\Omega\subset \mathbb R^n\, (n=2,\,3)$ be a heat conductor and $D$ a cavity with Robin boundary condition embedded in $\Omega$. We assume that the boundaries $\partial\Omega$ and $\partial D$ of $\Omega$ and $D$, respectively, are of class $C^2$. Injecting a heat flux $f$ on $\partial \Omega$ over some time interval $(0,\,T)$, the corresponding temperature distribution $u(x,\,t)$ in $\Omega\setminus\overline D\times(0,\,T)$ can be modeled by the following initial-boundary value problem:
\begin{equation}\label{eq:mp}
\left\{\begin{array}{l}
(\partial_t - \Delta)u=0\quad \textrm{ in } (\Omega\setminus\overline D)\times (0,\,T), \\
\partial_\nu u-\lambda u=0\quad\textrm{ on } \partial D\times (0,\,T), \\
\partial_\nu u=f \quad\textrm{ on } \partial \Omega\times (0,\,T),\\
u=0\quad\textrm{ at } t=0,
\end{array}\right.
\end{equation}
where $\lambda=\lambda(x)\in C^1(\partial D)$ is the real-valued impedance and $\nu$ on $\partial D$ (or $\partial \Omega$) is the unit normal vector directed into the exterior of $D$ (or $\Omega$). We will show that the initial-boundary value problem \eqref{eq:mp} is well-posed in a suitable Sobolev space. Then we idealize a set of many pairs of the heat flux $f$ and the corresponding surface temperature distribution $u|_{\partial D\times (0,\,T)}$ which is the measured data for active thermography as the Neumann-to-Dirichlet map $\Lambda_D$ given by $\Lambda_D:\,f \mapsto u|_{\partial D\times (0,\,T)}$. Thus, our inverse problem for thermography is to reconstruct $D$ from $\Lambda_D$.

When $D$ is a usual cavity with Neumann boundary condition or an inclusion, the corresponding inverse problem has been extensively studied. In \cite{B-C1997, D-R-V2006, D-V2010, E-I1997}, the uniqueness and stability estimate are established. In \cite{B-C2005, C-K-Y1998, C-K-Y1999}, Newton-type iteration algorithms based on domain derivatives are studied. As for non-iterative reconstruction schemes, we can consult the papers \cite{DKN,D-Y-L-N2009,I-K-N2010,KN,Y-K-N2010, Ikehata3, Ikehata4, I-P-S-T2012, N-S2013} and the references therein, where the reconstruction schemes called the dynamical probe method and the enclosure method are extensively studied. A reconstruction scheme for unknown cavities via Feynma-Kac type formula is also proposed in \cite{Kaw2015}. Recently, the authors established a linear sampling-type method for the heat equation to identify unknown cavities with Neumann boundary condition \cite{H-N-W2012}. This method was extended to the inclusion case in \cite{N-W2013}.

In this paper, we are concerned with the reconstruction of an unknown cavity where a Robin boundary condition is prescribed. Some results on uniqueness and stability for this inverse problem can be found in \cite{Bac2014, Isa2008}. Assuming that $D$ is given, the determination of the Robin coefficient $\lambda$ from measured data was also considered; see \cite{H-T-L2013, J-L2012} and the references therein. In this work, assuming that both the geometrical information of the cavity $D$ and the Robin coefficient $\lambda$ on its boundary are unknown, we establish a linear sampling method to identify the unknown Robin-type cavity $D$ from the boundary measurements $\Lambda_D$. As we know, the linear sampling method was originally proposed for inverse scattering problems in \cite{ C-K1996}, and was further investigated from both the theoretical and numerical aspects; for example, see \cite{A-L2009, C-C2006, C-C-M2011, K-G2008, LHY1, LHY2, LHY3, T-S1, T-S2}. Roughly speaking, this method in the heat equation case is based on the characterization of the approximate solvability of the so-called Neumann-to-Dirichlet map gap equation $(\Lambda_D-\Lambda_\emptyset)\Psi=\Gamma^0(x,\,t;\,y,\,s)$, where $\Lambda_\emptyset$ is the Neumann-to-Dirichlet map when there are not any cavities inside $\Omega$, and $\Gamma^0(x,\,t;\,y,\,s)$ is the Green function for the heat operator $\partial_t - \Delta$ in $\Omega\times(0,\,T)$ with homogeneous Neumann boundary condition on its boundary. By giving this characterization, we define a mathematical testing machine called an indicator function to reconstruct the boundary of $D$. When we probe it from the inside of $D$, we don't need to let the discrepancy of the Neumann-to-Dirichlet map gap equation tend to zero.

The new ingredients of this paper consists of two parts. First, we prove the well-posedness of the forward problem \eqref{eq:mp} in an anisotropic Sobolev space by the integral equation method. This is necessary for our purpose to investigate the inverse problem using layer potential argument, although the well-posedness can also be justified in the usual function space $W(0,\,T)$ by the argument in \cite{Wlo1987}. Second, compared with our previous works in \cite{H-N-W2012, N-W2013} for the Neumann cavity and inclusion cases, we provide a further investigation of the linear sampling method for the heat equation and show the asymptotic behavior of the indicator function by carefully analyzing that of the reflected solution of the fundamental solution. Since the reflected solution is the compensating term of the Green function for the related initial-boundary value problem, we actually give a pointwise short time asymptotic behavior of the Green function near $\partial D$. As a consequence, by observing the asymptotic behavior of the indicator function, we can have the followings:

(i) pointwise reconstruction of the shape and location of an unknown cavity;

(ii) information about the distance to the boundary of the cavity.
\\The more precise meaning of the above item (ii) is that we can know the distance to $\partial D$ when we probe it from inside $D$ by using the indicator function.

The rest of the paper is organized as follows. In Section \ref{FP}, we show the well-posedness of the forward problem in an anisotropic Sobolev space by the integral equation method. Then, in Section \ref{IP}, we present a mathematical justification of the linear sampling method, while the asymptotic behavior of the indicator function is provided in Section \ref{asy}. Finally, in Section \ref{con}, we give some concluding remarks.

\section{The integral equation method for the forward problem}\label{FP}
\setcounter{equation}{0}

In this section, we show the unique solvability of the forward problem by the integral equation method. Then we explicitly define the Neumann-to-Dirichlet map $\Lambda_D$ and state our inverse problem. Since our analysis for the inverse problem is based on the layer potential argument, it is necessary to establish the well-posedness of the forward problem in an anisotropic Sobolev space, instead of the usual function space $W(0,\,T)$.

Let us start by introducing the anisotropic Sobolev spaces. For $p,\,q\geq0$ we define
\begin{equation*}
H^{p,q}(\mathbb R^n\times\mathbb R):=L^2(\mathbb R;\,H^p(\mathbb R^n))\cap H^q(\mathbb R;\,L^2(\mathbb R^n)).
\end{equation*}
For $p,\,q\leq 0$ we define the space $H^{p,q}$ by duality $H^{p,q}(\mathbb R^n\times\mathbb R):=\left( H^{-p,-q}(\mathbb R^n\times\mathbb R)\right)^\prime$. Throughout this paper, we denote $X\times (0,\,T)$ and $\partial X\times (0,\,T)$ by $X_T$ and $(\partial X)_T$, respectively, where $X$ is a bounded domain in $\mathbb R^n$ and $\partial X$ denotes its boundary. By $H^{p,q}(X_T)$ we denote the space of restrictions of elements of $H^{p,q}(\mathbb R^n\times\mathbb R)$ to $X_T$. The space $H^{p,q}((\partial X)_T)$ is defined analogously. We also introduce the following function spaces:
\begin{eqnarray*}
&&\tilde H^{1,\frac{1}{2}}(X_T):=\left\{  u\in H^{1,\frac{1}{2}}(X\times (-\infty,\,T)) \big |\,u(x,\,t)=0\, \textrm{ for }t<0 \right\},\\
&&H^{1,\frac{1}{2}}(X_T;\,\partial_t-\Delta):=\left\{ u\in H^{1,\frac{1}{2}}(X_T) \big |\,(\partial_t-\Delta)u \in L^2(X_T)\right\}.
\end{eqnarray*}
Then our forward problem is formulated as follows.

\bigskip
{\bf Forward problem:} Given $f\in H^{-\frac{1}{2},-\frac{1}{4}}((\partial \Omega)_T)$ and $\lambda\in C^1(\partial D)$, find a unique solution $u\in \tilde H^{1,\frac{1}{2}}((\Omega\setminus\overline D)_T)$ to the problem \eqref{eq:mp}.

\bigskip
Denote by
$$\Gamma(x,\, t;\, y,\, s):=
\left\{ \begin{aligned}&\displaystyle\frac{1}{(4\pi (t-s))^{n/2}} \exp\left( -\frac{| x-y |^2}{4(t-s)} \right),&& t>s,\\
&0,&& t\leq s
\end{aligned}\right.
$$
the fundamental solution of the heat operator $\partial_t -\Delta$. For convenience, we sometimes write it as $\Gamma_{(y,\,s)}(x,\,t)$. Define the following heat layer potentials:
\begin{eqnarray*}
(V_{ij}\varphi)(x,\,t)&:=&\int_0^t\int_{S_i} \Gamma(x,\,t;\,y,\,s)\varphi(y,\,s)d\sigma(y)ds,\quad (x,\,t)\in S_j\times(0,\,T),\\
(K_{ij}\varphi)(x,\,t)&:=&\int_0^t\int_{S_i} \frac{\partial \Gamma(x,\,t;\,y,\,s)}{\partial \nu(y)}\varphi(y,\,s)d\sigma(y)ds,\quad (x,\,t)\in S_j\times(0,\,T),\\
(N_{ij}\varphi)(x,\,t)&:=&\int_0^t\int_{S_i} \frac{\partial \Gamma(x,\,t;\,y,\,s)}{\partial \nu(x)}\varphi(y,\,s)d\sigma(y)ds,\quad (x,\,t)\in S_j\times(0,\,T),\\
(W_{ij}\varphi)(x,\,t)&:=&-\frac{\partial}{\partial\nu(x)}\int_0^t\int_{S_i} \frac{\partial \Gamma(x,\,t;\,y,\,s)}{\partial \nu(y)}\varphi(y,\,s)d\sigma(y)ds,\quad (x,\,t)\in S_j\times(0,\,T).
\end{eqnarray*}
In this paper we take $i,\,j=1,\,2$ with $S_1=\partial D$ and $S_2=\partial \Omega$.

We now show the unique solvability of \eqref{eq:mp} in $\tilde H^{1,\frac{1}{2}}((\Omega\setminus\overline D)_T)$ by the integral equation method.

\begin{theorem}\label{th_forward}
Suppose that $\lambda\in C^1(\partial D)$ and $0\leq\lambda_1\leq
\lambda\leq\lambda_2$, where $\lambda_1$ and $\lambda_2$ are two given constants. Then, there exists a unique solution $u\in\tilde H^{1,\frac{1}{2}}((\Omega\setminus\overline D)_T)$ to the initial-boundary value problem \eqref{eq:mp} for any $f\in H^{-\frac{1}{2},-\frac{1}{4}}((\partial \Omega)_T)$.
\end{theorem}
{\bf Proof.} To prove the uniqueness of solutions in $\tilde H^{1,\frac{1}{2}}((\Omega\setminus\overline D)_T)$, we first establish the uniqueness in the function space $W(0,\,T)$ defined by
\begin{equation*}
W(0,\,T):=\left\{w:\;w\in L^2((0,\,T);\,H^1(\Omega\setminus\overline D)),\,\partial_t w \in L^2\left((0,\,T);\,\left(H^1(\Omega\setminus\overline D)\right)^\prime\right)\right\}.
\end{equation*}
Consider the bilinear form associated with \eqref{eq:mp}
\begin{equation*}
b(u,\,v,\,t):=\int_{\Omega\setminus\overline D} \nabla u \cdot \nabla v\, dx + \int_{\partial D} \lambda u v\, d\sigma(x)
\end{equation*}
for $u,\,v\in H^1(\Omega\setminus\overline D)$ and $t\in[0,\,T]$. It is easy to verify that
\begin{equation*}
|b(u,\,v,\,t)| \leq  c \|u\|_{H^1(\Omega\setminus\overline D)}\,\|v\|_{H^1(\Omega\setminus\overline D)},\quad
b(u,\,u,\,t) \geq  \|u\|^2_{H^1(\Omega\setminus\overline D)}- \| u \|^2_{L^2(\Omega\setminus\overline D)},
\end{equation*}
where $c$ is a positive constant. Then the uniqueness of solutions to \eqref{eq:mp} in $W(0,\,T)$ can be justified in a standard way; see, for example, \cite[\S 26]{Wlo1987}. Since any solution $u\in \tilde H^{1,\frac{1}{2}}((\Omega\setminus\overline D)_T)$ to $(\partial_t -\Delta)u=0$ with zero initial condition is in $W(0,\,T)$, we also have the uniqueness in $\tilde H^{1,\frac{1}{2}}((\Omega\setminus\overline D)_T)$.

\medskip
To prove the existence of the solution in $\tilde H^{1,\frac{1}{2}}((\Omega\setminus\overline D)_T)$, we make use of the integral equation method \cite{Costabel1990}. By Green's representation theorem for \eqref{eq:mp}, we have
\begin{eqnarray}\label{1.1}
u(x,\,t)&=&\int_0^t\int_{\partial(\Omega\setminus\overline D)} \Gamma(x,\,t;\,y,\,s)\frac{\partial u(y,\,s)}{\partial\nu(y)}d\sigma(y)ds - \int_0^t\int_{\partial(\Omega\setminus\overline D)}\frac{\partial \Gamma(x,\,t;\,y,\,s)}{\partial\nu(y)} u(y,\,s)d\sigma(y)ds \nonumber\\
&=&-\int_0^t\int_{\partial D} \Gamma(x,\,t;\,y,\,s)\frac{\partial u(y,\,s)}{\partial\nu(y)}d\sigma(y)ds + \int_0^t\int_{\partial D}\frac{\partial\Gamma(x,\,t;\,y,\,s)}{\partial\nu(y)} u(y,\,s)d\sigma(y)ds\nonumber\\
&&+\int_0^t\int_{\partial\Omega} \Gamma(x,\,t;\,y,\,s)\frac{\partial u(y,\,s)}{\partial\nu(y)}d\sigma(y)ds - \int_0^t\int_{\partial\Omega}\frac{\partial \Gamma(x,\,t;\,y,\,s)}{\partial\nu(y)} u(y,\,s)d\sigma(y)ds.
\end{eqnarray}
Set $\left(u_1,\,u_2,\,u_3\right)^T=\left(\partial_\nu u|_{(\partial D)_T},\,u|_{(\partial D)_T},\,u|_{(\partial \Omega)_T} \right)^T$. Then, using the jump relations of layer potentials, we deduce the following system of boundary integral equations:
\begin{equation}\label{1.2}
\left(
\begin{array}{ccc}
V_{11} & \frac{1}{2}I-K_{11} & K_{21}\\
-\frac{1}{2}I + N_{11}\; &  W_{11}+\lambda I & -W_{21}\\
V_{12} & -K_{12} & \frac{1}{2}I + K_{22}
\end{array}
\right) \,
\left(
\begin{array}{c}
u_1\\
u_2\\
u_3
\end{array}
\right)=
\left(
\begin{array}{c}
V_{21}f\\
N_{21}f\\
V_{22}f
\end{array}
\right).
\end{equation}

Define the space $\mathcal H:=H^{-\frac{1}{2},-\frac{1}{4}}((\partial D)_T)\times H^{\frac{1}{2},\frac{1}{4}}((\partial D)_T)\times H^{\frac{1}{2},\frac{1}{4}}((\partial \Omega)_T)$ with the norm
\begin{equation*}
\|(u_1,\,u_2,\,u_3)^T\|_{\mathcal H}:=\left( \| u_1 \|^2_{H^{-\frac{1}{2},-\frac{1}{4}}((\partial D)_T)} +  \| u_2 \|^2_{H^{\frac{1}{2},\frac{1}{4}}((\partial D)_T)}+\| u_3 \|^2_{H^{\frac{1}{2},\frac{1}{4}}((\partial \Omega)_T)}\right)^{1/2}
\end{equation*}
for $(u_1,\,u_2,\,u_3)^T\in\mathcal H$. We show the unique solvability of the system \eqref{1.2} in $\mathcal H$. To accomplish this, we define the operator
\begin{equation}\label{1.3}
\mathcal A:=\left(
\begin{array}{cc}
V_{11} & \frac{1}{2}I-K_{11}\\
-\frac{1}{2}I + N_{11}\; &  W_{11}+\lambda I
\end{array}
\right)
\end{equation}
and the space $\mathcal H_0:=H^{-\frac{1}{2},-\frac{1}{4}}((\partial D)_T)\times H^{\frac{1}{2},\frac{1}{4}}((\partial D)_T)$ equipped with the norm
\begin{equation*}
\|(u_1,\,u_2)^T\|_{\mathcal H_0}:=\left( \| u_1 \|^2_{H^{-\frac{1}{2},-\frac{1}{4}}((\partial D)_T)} +  \| u_2 \|^2_{H^{\frac{1}{2},\frac{1}{4}}((\partial D)_T)}\right)^{1/2}  \quad \textrm{ for }\,(u_1,\,u_2)^T\in\mathcal H_0.
\end{equation*}
Let $\mathcal H_0^\prime:=H^{\frac{1}{2},\frac{1}{4}}((\partial D)_T)\times H^{-\frac{1}{2},-\frac{1}{4}}((\partial D)_T)$ be the dual space of $\mathcal H_0$ with the duality between $\mathcal H_0$ and $\mathcal H_0^\prime$ defined by
\begin{equation*}
\left\langle
\left(
\begin{array}{c}
\varphi_1\\
\varphi_2
\end{array}
\right),
\left(
\begin{array}{c}
\psi_1\\
\psi_2
\end{array}
\right)
\right\rangle=\langle \varphi_1, \, \psi_1 \rangle + \langle \psi_2,\,\varphi_2 \rangle \quad \mathrm{for}\; (\varphi_1,\,\varphi_2)^T\in\mathcal H_0^\prime,\, (\psi_1,\,\psi_2)^T\in\mathcal H_0.
\end{equation*}

In the following, we first prove the invertibility of $\mathcal A:\,\mathcal H_0 \to \mathcal H_0^\prime$. Define a bilinear form $a(\cdot,\,\cdot)$ on $\mathcal H_0\times \mathcal H_0$ associated with the operator $\mathcal A$ by
\begin{eqnarray*}
a\left((u_1,\,u_2)^T,\,(v_1,\,v_2)^T\right)&:=&\left\langle  \mathcal A\left( \begin{array}{c} u_1 \\ u_2 \end{array}\right),\, \left( \begin{array}{c} v_1 \\ v_2 \end{array} \right)\right\rangle\\
&=&\langle V_{11}u_1,\,v_1\rangle+\langle (\frac{1}{2}I-K_{11})u_2,\,v_1 \rangle +\langle v_2,\,(-\frac{1}{2}I+N_{11})u_1 \rangle\\
&& + \langle v_2,\, W_{11}u_2\rangle +\langle v_2,\, \lambda u_2\rangle \quad \textrm{ for }(u_1,\,u_2)^T,\, (v_1,\,v_2)^T\in\mathcal H_0.
\end{eqnarray*}
There exists a positive constant $C_1$ such that
\begin{equation}\label{co1}
|a\left((u_1,\,u_2)^T,\,(v_1,\,v_2)^T\right)|\leq C_1 \| (u_1,\,u_2)^T \|_{\mathcal H_0}\,\|(v_1,\,v_2)^T\|_{\mathcal H_0}.
\end{equation}
Note that
\begin{eqnarray*}
&&a\left((u_1,\,u_2)^T,\,(u_1,\,u_2)^T\right)\\
&=&\langle V_{11}u_1,\,u_1\rangle+\langle -K_{11}u_2,\,u_1 \rangle +\langle u_2,\,+N_{11}u_1 \rangle + \langle u_2,\, W_{11}u_2\rangle +\langle u_2,\, \lambda u_2\rangle\\
&=&\left\langle  \left(
\begin{array}{cc}
V_{11} & -K_{11}\\
N_{11} &  W_{11}
\end{array}
\right)\left( \begin{array}{c} u_1 \\ u_2 \end{array}\right),\, \left( \begin{array}{c} u_1 \\ u_2 \end{array} \right)\right\rangle + \langle u_2,\, \lambda u_2\rangle,
\end{eqnarray*}
where the operator
\begin{equation*}
\mathcal A_0:=\left(
\begin{array}{cc}
V_{11} & -K_{11}\\
N_{11} &  W_{11}
\end{array}
\right)
\end{equation*}
is positive on $\mathcal H_0$ (see \cite[Theorem 3.11]{Costabel1990}). Then we have the estimate
\begin{equation}\label{co2}
a\left((u_1,\,u_2)^T,\,(u_1,\,u_2)^T\right)\geq C_2\| (u_1,\,u_2)^T \|^2_{\mathcal H_0},
\end{equation}
where $C_2$ is a positive constant. From \eqref{co1} and \eqref{co2} we conclude that $\mathcal A:\,\mathcal H_0\to \mathcal H_0^\prime$ is an isomorphism.

In addition, it follows from Corollary 3.14 in \cite{Costabel1990} that
\begin{equation*}
\frac{1}{2}I+K_{22}:\,H^{\frac{1}{2},\frac{1}{4}}((\partial \Omega)_T)\to H^{\frac{1}{2},\frac{1}{4}}((\partial \Omega)_T)
\end{equation*}
is also an isomorphism. Since $\Gamma(x,\,t;\,y,\,s)$ is smooth enough for $x\not=y,\,0\leq s \leq t\leq T$, the following operators
\begin{equation*}
\begin{array}{lcl}
K_{21}:\,H^{\frac{1}{2},\frac{1}{4}}((\partial \Omega)_T)\to H^{\frac{1}{2},\frac{1}{4}}((\partial D)_T), & &W_{21}:\,H^{\frac{1}{2},\frac{1}{4}}((\partial \Omega)_T)\to H^{-\frac{1}{2},-\frac{1}{4}}((\partial D)_T),\\
V_{12}:\,H^{-\frac{1}{2},-\frac{1}{4}}((\partial D)_T)\to H^{\frac{1}{2},\frac{1}{4}}((\partial \Omega)_T), & &K_{12}:\,H^{\frac{1}{2},\frac{1}{4}}((\partial D)_T)\to H^{\frac{1}{2},\frac{1}{4}}((\partial \Omega)_T)
\end{array}
\end{equation*}
are compact. Therefore, the system \eqref{1.2} is Fredholm with index zero. To show the unique solvability of \eqref{1.2}, it suffices to show its uniqueness.

Let $(\varphi_1,\,\varphi_2,\,\varphi_3)^T\in \mathcal H$ be the solution to the homogeneous form of \eqref{1.2}. That is,
\begin{equation}\label{1.2_0}
\left(
\begin{array}{ccc}
V_{11} & \frac{1}{2}I-K_{11} & K_{21}\\
-\frac{1}{2}I + N_{11}\; &  W_{11}+\lambda I & -W_{21}\\
V_{12} & -K_{12} & \frac{1}{2}I + K_{22}
\end{array}
\right) \,
\left(
\begin{array}{c}
\varphi_1\\
\varphi_2\\
\varphi_3
\end{array}
\right)=
\left(
\begin{array}{c}
0\\
0\\
0
\end{array}
\right).
\end{equation}
Define
\begin{eqnarray}\label{1.4}
w(x,\,t)&:=&-\int_0^t\int_{\partial D} \Gamma(x,\,t;\,y,\,s)\varphi_1(y,\,s)d\sigma(y)ds + \int_0^t\int_{\partial D}\frac{\partial\Gamma(x,\,t;\,y,\,s)}{\partial\nu(y)} \varphi_2(y,\,s)d\sigma(y)ds\nonumber\\
&&- \int_0^t\int_{\partial\Omega}\frac{\partial \Gamma(x,\,t;\,y,\,s)}{\partial\nu(y)} \varphi_3(y,\,s)d\sigma(y)ds, \quad (x,\,t)\in\left( \mathbb R^n\setminus(\partial D\cup \partial\Omega) \right)_T.
\end{eqnarray}
Note that $w(x,\,t)$ satisfies
\begin{equation}\label{1.5}
\partial_t w- \Delta w=0 \quad \textrm{ in } (\mathbb R^n\setminus\overline \Omega)_T
\end{equation}
and
\begin{equation}\label{1.6}
w=0 \quad \textrm{ at } t=0.
\end{equation}
Letting $x$ tend to the boundary $\partial\Omega$, we derive from the jump relations of layer potentials that
\begin{equation*}
w^+=-V_{12}\varphi_1 + K_{12}\varphi_2 - (\frac{1}{2}I + K_{22})\varphi_3.
\end{equation*}
Throughout this paper we use \lq$+$\rq{} and \lq$-$\rq{} to denote the limits taken from the exterior and interior of a domain, respectively. It implies from the third equation of \eqref{1.2_0} that
\begin{equation}\label{1.7}
w^+=0 \quad  \textrm{ on } (\partial\Omega)_T.
\end{equation}
Then, using the same argument as in \cite[Appendix A]{H-N-W2012}, we can prove that $w=0$ in $(\mathbb R^n\setminus\overline \Omega)_T$. It follows that $\partial_\nu w^+=0$ on $(\partial\Omega)_T$, and hence $\partial_\nu w^-=0$ on $(\partial\Omega)_T$ due to the jump relations of layer potentials.

On the other hand, the function $w$ defined by \eqref{1.4} also satisfies the heat equation in $(\Omega\setminus\overline D)_T$. Using the jump relations of layer potentials again, we derive that
\begin{eqnarray*}
w&=&-V_{11}\varphi_1 + K_{11}\varphi_2 + \frac{1}{2}\varphi_2 - K_{21}\varphi_3 \quad  \textrm{ on } (\partial D)_T,\\
\partial_\nu w&=&\frac{1}{2}\varphi_1-N_{11}\varphi_1-W_{11}\varphi_2 +W_{21}\varphi_3 \quad  \textrm{ on } (\partial D)_T.
\end{eqnarray*}
In terms of the first two equations of \eqref{1.2_0}, we have $w=\varphi_2$ and $\partial_\nu w=\lambda\varphi_2$ on $(\partial D)_T$.
In conclusion, the function $w$ defined by \eqref{1.4} satisfies
\begin{equation}\label{eq:mp_0}
\left\{\begin{array}{l}
(\partial_t - \Delta)w=0\quad \textrm{ in } (\Omega\setminus\overline D)_T, \\
\partial_\nu w-\lambda w=0\quad\textrm{ on } (\partial D)_T, \\
\partial_\nu w=0 \quad\textrm{ on } (\partial \Omega)_T,\\
w=0\quad\textrm{ at } t=0.
\end{array}\right.
\end{equation}
By the uniqueness of solutions to \eqref{eq:mp}, we have $w=0$ in $(\Omega\setminus\overline D)_T$, and hence
\begin{equation}\label{1.9}
w^-=0\quad \textrm{ on } (\partial\Omega)_T.
\end{equation}
Combining \eqref{1.7} and \eqref{1.9}, we obtain $\varphi_3=w^--w^+=0$ on $(\partial\Omega)_T$.
As a consequence, we have
\begin{equation*}
\left(
\begin{array}{cc}
V_{11} & \frac{1}{2}I-K_{11}\\
-\frac{1}{2}I + N_{11}\; &  W_{11}+\lambda I
\end{array}
\right) \,
\left(
\begin{array}{c}
\varphi_1\\
\varphi_2
\end{array}
\right)=
\left(
\begin{array}{c}
0\\
0
\end{array}
\right).
\end{equation*}
Now by the invertibility of $\mathcal A$, we conclude that $\varphi_1=\varphi_2=0$ on $(\partial D)_T$. Due to the Fredholm theory, the unique solvability of the system \eqref{1.2} is justified. Moreover, according to the mapping properties of heat layer potentials, the function $u$ expressed by \eqref{1.1} is the desired solution in $\tilde H^{1,\frac{1}{2}}((\Omega\setminus\overline D)_T)$. The proof is complete. \hfill $\Box$

\bigskip
Based on our Theorem \ref{th_forward} and Lemma 2.4 in \cite{Costabel1990}, we can define the Neumann-to-Dirichlet map $\Lambda_D$ by
\begin{equation}\label{eq:ND}
\Lambda_D:\, H^{-\frac{1}{2},-\frac{1}{4}}((\partial \Omega)_T)\to H^{\frac{1}{2},\frac{1}{4}}((\partial \Omega)_T),\quad f\mapsto u^f|_{(\partial\Omega)_T}.
\end{equation}
If there is not any cavity inside $\Omega$, i.e., $D=\emptyset$, the unique solvability of the forward problem in $\tilde H^{1,\frac{1}{2}}(\Omega_T)$ can be found in \cite{Costabel1990}. In this case, we denote the Neumann-to-Dirichlet map by $\Lambda_\emptyset$. Taking the Neumann-to-Dirichlet map $\Lambda_D$ as the measured data, our inverse problem for \eqref{eq:mp} is formulated as follows:

{\bf Inverse Problem:} Reconstruct $D$ from $\Lambda_D$.

\section{The linear sampling method for the inverse problem}\label{IP}
\setcounter{equation}{0}

In this section, we present a linear sampling-type method for the inverse problem formulated above and give its mathematical justification by using the layer potential argument. Let $\Gamma^0_{(y,\,s)}(x,\,t):=\Gamma^0(x,\,t;\,y,\,s)$ be the Green function of the heat operator in $\Omega_T$ with Neumann boundary condition on its boundary $(\partial\Omega)_T$. Then the linear sampling method for the heat equation is based on the characterization of the approximate solution to the Neumann-to-Dirichlet map gap equation
\begin{equation}\label{gap}
(\Lambda_D - \Lambda_\emptyset) g = \Gamma^0_{(y,\,s)}(x,\,t), \quad (x,\,t)\in (\partial\Omega)_T,
\end{equation}
where $s \in (0,\,T)$ is a fixed time and $y\in \Omega$ is the sampling point. By this characterization, we can define an indicator function as a mathematical testing machine to reconstruct the location and shape of $D$.

To begin with, let us define the operators $S,\,H,\,A$ and $F$ as follows.
\begin{itemize}
\item Define $S:\,H^{-\frac{1}{2},-\frac{1}{4}}((\partial \Omega)_T)\to \tilde H^{1,\frac{1}{2}}(D_T)$ and $H:\,H^{-\frac{1}{2},-\frac{1}{4}}((\partial \Omega)_T)\to H^{-\frac{1}{2},-\frac{1}{4}}((\partial D)_T)$ by
\begin{equation}\label{S}
S:\,f\mapsto u^f|_{D_T},
\end{equation}
and
\begin{equation}\label{H}
H:\,f\mapsto (\partial_\nu u^f-\lambda u^f)|_{(\partial D)_T},
\end{equation}
respectively, where $u^f$ is the solution to
\begin{equation}\label{eq:S}
\left\{ \begin{array}{l}
(\partial_t - \Delta)u^f=0 \quad \textrm{ in }\Omega_T,\\
\partial_\nu u^f =f \quad \textrm{ on }(\partial\Omega)_T,\\
u^f=0 \quad \textrm{ at } t=0.
\end{array}\right.
\end{equation}

\item Define $A:\,H^{-\frac{1}{2},-\frac{1}{4}}((\partial D)_T) \to H^{\frac{1}{2},\frac{1}{4}}((\partial \Omega)_T)$ by
\begin{equation}\label{A}
A:\, g\mapsto z^g|_{(\partial\Omega)_T},
\end{equation}
where $z^g$ is the solution to
\begin{equation}\label{eq:A}
\left\{ \begin{array}{l}
(\partial_t - \Delta)z^g=0 \quad \textrm{ in }(\Omega\setminus\overline D)_T,\\
\partial_\nu z^g-\lambda z^g=g \quad  \textrm{ on }(\partial D)_T,\\
\partial_\nu z^g=0  \quad \textrm{ on }(\partial \Omega)_T,\\
z^g=0 \quad  \textrm{ at } t=0.
\end{array}\right.
\end{equation}

\item Define $F:\,H^{-\frac{1}{2},-\frac{1}{4}}((\partial \Omega)_T) \to H^{\frac{1}{2},\frac{1}{4}}((\partial \Omega)_T)$ by
$F:=\Lambda_D - \Lambda_{\emptyset}$.
\end{itemize}

To characterize the solution to \eqref{gap}, we first investigate the operator $F$ and prove the following lemmas.

\begin{lemma}\label{lem:1}
The operator $F$ can be factorized as $F=-AH$.
\end{lemma}
{\bf Proof.} For $f\in H^{-\frac{1}{2},-\frac{1}{4}}((\partial \Omega)_T)$, let $u$ and $v$ be such that
\begin{equation*}
\left\{ \begin{array}{l}
(\partial_t - \Delta)u=0 \quad\textrm{ in }(\Omega\setminus\overline D)_T,\\
\partial_\nu u - \lambda u=0 \quad \textrm{ on }(\partial D)_T,\\
\partial_\nu u= f \quad \textrm{ on }(\partial \Omega)_T,\\
u=0 \quad \textrm{ at } t=0,
\end{array}\right.
\end{equation*}
and
\begin{equation*}
\left\{ \begin{array}{l}
(\partial_t - \Delta)v=0 \quad \textrm{ in }\Omega_T,\\
\partial_\nu v =f \quad \textrm{ on }(\partial\Omega)_T,\\
v=0 \quad \textrm{ at } t=0,
\end{array}\right.
\end{equation*}
respectively. Define $w$ as the solution to
\begin{equation*}
\left\{ \begin{array}{l}
(\partial_t - \Delta)w=0 \quad \textrm{ in }(\Omega\setminus\overline D)_T,\\
\partial_\nu w-\lambda w=-(\partial_\nu v-\lambda v) \quad \textrm{ on }(\partial D)_T,\\
\partial_\nu w=0 \quad \textrm{ on }(\partial \Omega)_T,\\
w=0 \quad \textrm{ at } t=0.
\end{array}\right.
\end{equation*}
Then it holds that
\begin{equation*}
\left\{ \begin{array}{l}
(\partial_t - \Delta)(u-v-w)=0 \quad \textrm{ in }(\Omega\setminus\overline D)_T,\\
\partial_\nu (u-v-w)-\lambda (u-v-w)=0 \quad \textrm{ on }(\partial D)_T,\\
\partial_\nu (u-v-w)=0 \quad \textrm{ on }(\partial \Omega)_T,\\
u-v-w=0 \quad \textrm{ at } t=0.
\end{array}\right.
\end{equation*}
The uniqueness result in Theorem \ref{th_forward} says that $u-v=w$ in $\tilde H^{1,\frac{1}{2}}((\Omega\setminus\overline D)_T)$. Hence, we have
\begin{equation*}
A(-Hf)=w|_{(\partial\Omega)_T}=(u-v)|_{(\partial \Omega)_T}=(\Lambda_D - \Lambda_\emptyset)f=Ff,
\end{equation*}
which completes the proof. \hfill $\Box$

\begin{lemma}\label{lem:2}
The operator $H:\,H^{-\frac{1}{2},-\frac{1}{4}}((\partial \Omega)_T)\to H^{-\frac{1}{2},-\frac{1}{4}}((\partial D)_T)$ is continuous and has a dense range.
\end{lemma}
{\bf Proof.} We prove the result using the layer potential argument. For $f\in H^{-\frac{1}{2},-\frac{1}{4}}((\partial \Omega)_T)$, we consider the following initial-boundary value problem:
\begin{equation}\label{eq:den1}
\left\{ \begin{array}{l}
(\partial_t - \Delta)u=0 \quad \textrm{ in }\Omega_T,\\
\partial_\nu u =f \quad \textrm{ on }(\partial\Omega)_T,\\
u=0 \quad \textrm{ at } t=0.
\end{array}\right.
\end{equation}
Express its solution by a single-layer heat potential
\begin{equation}\label{eq:den2}
u(x,\,t)=V_0\psi:=\int_0^t \int_{\partial \Omega} \Gamma(x,\,t;\,y,\,s) \psi(y,\,s)d\sigma(y)ds,\quad (x,\,t)\in \Omega_T
\end{equation}
with an unknown density $\psi\in H^{-\frac{1}{2},-\frac{1}{4}}((\partial\Omega)_T)$. Using the jump relations of layer potentials, the problem \eqref{eq:den1} is reformulated as the boundary integral equation
\begin{equation}\label{eq:den3}
(\frac{1}{2}I+N_{22})\psi = f \quad \textrm{ on }(\partial\Omega)_T,
\end{equation}
where the operator $\frac{1}{2}I+N_{22}:\,H^{-\frac{1}{2},-\frac{1}{4}}((\partial\Omega)_T) \to H^{-\frac{1}{2},-\frac{1}{4}}((\partial\Omega)_T)$
is an isomorphism (see \cite[Corollary 3.14]{Costabel1990}).

Define $\tilde V\psi:=\left(\partial_\nu (V_0\psi) -\lambda (V_0\psi)\right) |_{(\partial D)_T}$. Then the operator $H$ can be represented as
$$H=\tilde V(\frac{1}{2}I+N_{22})^{-1}.$$
Thus, the continuity of $H$ is evident. To prove the denseness property of $H$, it suffices to show that the operator $\tilde V:\,H^{-\frac{1}{2},-\frac{1}{4}}((\partial\Omega)_T)\to H^{-\frac{1}{2},-\frac{1}{4}}((\partial D)_T)$ has a dense range.

Indeed, by direct calculations, we have
\begin{equation}\label{eq:den6}
\tilde V\psi(x,\,t)=\int_0^t \int_{\partial\Omega}M(x,\,t;\,y,\,s)\psi(y,\,s)d\sigma(y)ds,\quad (x,\,t)\in (\partial D)_T
\end{equation}
with $ M(x,\,t;\,y,\,s):=\partial_{\nu(x)} \Gamma(x,\,t;\,y,\,s)-\lambda(x)\, \Gamma(x,\,t;\,y,\,s)$. Denote by $\tilde V^*:\,H^{\frac{1}{2},\frac{1}{4}}((\partial D)_T)\to H^{\frac{1}{2},\frac{1}{4}}((\partial\Omega)_T)$ the transpose of $\tilde V$ in the sense that $ \langle\tilde V \psi,\,\eta\rangle=\langle\psi,\,\tilde V^*\eta\rangle $ for any $\psi\in H^{-\frac{1}{2},-\frac{1}{4}}((\partial\Omega)_T)$ and $\eta\in H^{\frac{1}{2},\frac{1}{4}}((\partial D)_T)$. It follows that
\begin{equation}\label{eq:den7}
\tilde V^*\eta(y,\,s)=\int_s^T \int_{\partial D}M(x,\,t;\,y,\,s)\eta(x,\,t)d\sigma(x)dt,\quad (y,\,s)\in (\partial \Omega)_T.
\end{equation}
Then, to show that the range of $\tilde V$ is dense, we are led to prove the injectivity of $\tilde V^*$, that is, $\eta=0$ in $H^{\frac{1}{2},\frac{1}{4}}((\partial D)_T)$ if we have $ \tilde V^*\eta=0$ on $(\partial\Omega)_T$. To this end, we define
\begin{equation}\label{eq:den8}
w(y,\,s)=\int_s^T \int_{\partial D}M(x,\,t;\,y,\,s)\eta(x,\,t)d\sigma(x)dt,\quad (y,\,s)\in (\mathbb{R}^n\setminus\partial D)_T.
\end{equation}
Noticing that $w$ satisfies
\begin{equation*}
\left\{\begin{array}{l}
\partial_s w + \Delta w=0 \quad \textrm{ in } (\mathbb{R}^n\setminus\overline\Omega)_T,\\
w=0 \quad\textrm{ on }(\partial\Omega)_T,\\
w=0 \quad\textrm{ at }s=T,
\end{array}\right.
\end{equation*}
we have $w=0$ in $(\mathbb{R}^n\setminus\overline\Omega)_T$ by the same argument as in \cite[Appendix A]{H-N-W2012}. Using the unique continuation principle, we further have
\begin{equation}\label{add1}
w=0\quad \textrm{ in } (\mathbb{R}^n\setminus\overline{D})_T,
\end{equation}
and therefore
\begin{equation}\label{add2}
\frac{\partial w^+}{\partial\nu}=0,\quad w^+=0 \quad \textrm{ on } (\partial D)_T.
\end{equation}

Note that
\begin{eqnarray*}
w(y,\,s)&=&\int_s^T \int_{\partial D}M(x,\,t;\,y,\,s)\eta(x,\,t)d\sigma(x)dt\\
&=&\int_s^T \int_{\partial D}\left(\partial_{\nu(x)} \Gamma(x,\,t;\,y,\,s)-\lambda(x) \Gamma(x,\,t;\,y,\,s)\right)\eta(x,\,t)d\sigma(x)dt\\
&=&\int_0^{T-s}\int_{\partial D}\left(\partial_{\nu(x)} \Gamma(y,\,T-s;\,x,\,\tau)-\lambda(x) \Gamma(y,\,T-s;\,x,\,\tau)\right)\tilde\eta(x,\,\tau)d\sigma(x)d\tau,
\end{eqnarray*}
where $\tilde \eta(x,\,\tau)=\eta(x,\,T-\tau)$. Using the jump relations of layer potentials, we have
\begin{equation}\label{add4}
\frac{\partial w^+}{\partial\nu}-\frac{\partial w^-}{\partial\nu}=\lambda\eta,\quad w^+-w^-=\eta \quad \textrm{ on } (\partial D)_T.
\end{equation}
It implies from \eqref{add2} and \eqref{add4} that
\begin{equation}\label{add5}
\frac{\partial w^-}{\partial\nu}-\lambda w^-=0 \quad \textrm{ on } (\partial D)_T.
\end{equation}
Observe that $w$ also meets
\begin{equation*}
\left\{\begin{array}{l}
\partial_s w + \Delta w=0 \quad \textrm{ in } D_T,\\
w=0 \quad\textrm{ at }s=T.
\end{array}\right.
\end{equation*}
Then, by the uniqueness of solutions to the backward problem, we obtain that
\begin{equation}\label{add6}
w=0\quad \textrm{ in } D_T.
\end{equation}
Thus, it can be concluded from \eqref{add1}, \eqref{add4} and \eqref{add6} that $\eta=0$. This completes the proof. \hfill  $\Box$

\begin{lemma}\label{lem:4}
The operator $A:\,H^{-\frac{1}{2},-\frac{1}{4}}((\partial D)_T) \to H^{\frac{1}{2},\frac{1}{4}}((\partial \Omega)_T)$ is injective, compact and has a dense range.
\end{lemma}
{\bf Proof.} The injectivity can be easily seen from the unique continuation principle for the heat operator $\partial_t - \Delta$. We now show the denseness. Let $g_j\in H^{-\frac{1}{2},-\frac{1}{4}}((\partial D)_T)\,(j\in\mathbb{N})$ be such that the linear hull of $\{g_j\}$ is dense in $H^{-\frac{1}{2},-\frac{1}{4}}((\partial D)_T)$. Then it is enough to prove $f=0$ if
\begin{equation*}
\int_{(\partial \Omega)_T}\varphi_j f d\sigma(x)dt =0
\end{equation*}
for $f\in H^{-\frac{1}{2},-\frac{1}{4}}((\partial \Omega)_T)$ and all $\varphi_j:=Ag^j=z^{g_j}|_{(\partial \Omega)_T}$. Here we recall that $z^{g_j}$ is the solution to \eqref{eq:A} with $g=g_j$.

Let $v\in\tilde H^{1,\frac{1}{2}}((\Omega\setminus\overline D)_T)$ be the solution to the following backward problem
\begin{equation*}
\left\{ \begin{array}{l}
(\partial_t + \Delta)v=0 \quad\textrm{ in }(\Omega\setminus\overline D)_T,\\
\partial_\nu v - \lambda v=0 \quad \textrm{ on }(\partial D)_T,\\
\partial_\nu v= f \quad \textrm{ on }(\partial \Omega)_T,\\
v=0 \quad \textrm{ at }t=T,
\end{array}\right.
\end{equation*}
and set $z_j=z^{g_j}$. Then, we have
\begin{eqnarray*}
0&=&\int_{(\Omega\setminus\overline D)_T} ( v \Delta z_j - z_j \Delta v )dxdt \\
&=& \int_{(\partial\Omega)_T} ( \partial_\nu z_j \, v - \partial_\nu v\, z_j )d\sigma(x)dt - \int_{(\partial D)_T} ( \partial_\nu z_j\, v - \partial_\nu v\, z_j )d\sigma(x)dt \\
&=& -\int_{(\partial D)_T} g_j v d\sigma(x)dt.
\end{eqnarray*}
So $v=0$ on $(\partial D)_T$. By the boundary condition of $v$, we further have $\partial_\nu v|_{(\partial D)_T}=0$. Therefore, $v=0$ in $(\Omega\setminus\overline D)_T$, and then $f=\partial_\nu v|_{(\partial \Omega)_T}=0$.

Finally, let us prove the compactness of $A$. Indeed, there exists a unique density $\varphi\in H^{-\frac{1}{2},-\frac{1}{4}}((\partial D)_T)$ such that the solution $z^g$ to (\ref{eq:A}) with $g\in H^{-\frac{1}{2},-\frac{1}{4}}((\partial D)_T)$ is given by
\begin{equation*}
z^g(x,\,t)=\int_0^t\int_{\partial D} \Gamma^0(x,\,t;\,y,\,s)\varphi(y,\,s)d\sigma(y)ds.
\end{equation*}
Notice that $\Gamma^0(x,\,t;\,y,\,s)$ is smooth enough for $x\in\partial\Omega,\,y\in\partial D$ and $0\leq s\leq t\leq T$. It can be concluded that $Ag=z^g|_{(\partial\Omega)_T}\in C^\infty(\partial\Omega\times[0,\,T])$, and hence $A$ is compact. This  completes the proof. \hfill $\Box$

\bigskip
We are now in a position to state our main results, which motivate the linear sampling method for reconstructing $D$.
\begin{theorem}\label{prop:3}
Let $s\in(0,\,T)$ be fixed. For $y\in D$, there exists a function $g^y\in H^{-\frac{1}{2},-\frac{1}{4}}((\partial\Omega)_T)$ satisfying
\begin{equation}\label{eq:data1}
\| Fg^y - \Gamma^0_{(y,\,s)} \|_{H^{\frac{1}{2},\frac{1}{4}}((\partial\Omega)_T)}< \varepsilon
\end{equation}
such that
\begin{equation}\label{eq:density11}
\lim_{y\to\partial D} \| g^y \|_{H^{-\frac{1}{2},-\frac{1}{4}}((\partial\Omega)_T)}=\infty
\end{equation}
and
\begin{equation}\label{eq:density12}
\lim_{y\to\partial D} \| Sg^y \|_{\tilde H^{1,\frac{1}{2}}(D_T)}=\infty.
\end{equation}
\end{theorem}
{\bf Proof.} According to Lemma \ref{lem:2}, for any $\varepsilon > 0$ there exists a function $g^y\in H^{-\frac{1}{2},-\frac{1}{4}}((\partial\Omega)_T)$ such that
\begin{equation}\label{badd0}
\left\| Hg^y + \left(\partial_\nu \Gamma^0_{(y,\,s)} - \lambda \Gamma^0_{(y,\,s)}\right) \right\|_{H^{-\frac{1}{2},-\frac{1}{4}}((\partial D)_T)}< \frac{\varepsilon}{\| A \|},
\end{equation}
where $\| A \|$ is the norm of the operator $A$ defined by \eqref{A}. Since $\Gamma^0_{(y,\,s)}$ satisfies the heat equation in $(\Omega\setminus\overline D)_T$ and $\partial_\nu \Gamma^0_{(y,\,s)}|_{(\partial\Omega)_T}=0$, it holds that
$$
A\left( \partial_\nu \Gamma^0_{(y,\,s)}-\lambda \Gamma^0_{(y,\,s)} \right)=\Gamma^0_{(y,\,s)}|_{(\partial\Omega)_T}.
$$
By Lemma \ref{lem:1}, we have
\begin{eqnarray*}
\left\| F g^y - \Gamma^0_{(y,\,s)} \right\| _{H^{\frac{1}{2},\frac{1}{4}}((\partial\Omega)_T)}
&=& \left\| -AHg^y - A\left( \partial_\nu \Gamma^0_{(y,\,s)}-\lambda \Gamma^0_{(y,\,s)} \right) \right\|_{H^{\frac{1}{2},\frac{1}{4}}((\partial\Omega)_T)}\\
&\leq & \| A \| \, \left\| H g^y +\left(\partial_\nu \Gamma^0_{(y,\,s)} -\lambda \Gamma^0_{(y,\,s)}\right) \right\|_{H^{-\frac{1}{2},-\frac{1}{4}}((\partial D)_T)} < \varepsilon.
\end{eqnarray*}
Hence, due to the boundedness of $H$, we have
\begin{eqnarray}\label{badd1}
&&\| g^y \|_{H^{-\frac{1}{2},-\frac{1}{4}}((\partial\Omega)_T)} \geq  C\| H g^y \|_{H^{-\frac{1}{2},-\frac{1}{4}}((\partial D)_T)}\nonumber\\
&\geq & C \left( \left\| \partial_\nu \Gamma^0_{(y,\,s)} -\lambda \Gamma^0_{(y,\,s)} \right\|_{H^{-\frac{1}{2},-\frac{1}{4}}((\partial D)_T)}- \left\| H g^y + \left(\partial_\nu \Gamma^0_{(y,\,s)} -\lambda \Gamma^0_{(y,\,s)}\right)\right \|_{H^{-\frac{1}{2},-\frac{1}{4}}((\partial D)_T)} \right)\nonumber\\
&\geq & C \left \| \partial_\nu \Gamma^0_{(y,\,s)} -\lambda \Gamma^0_{(y,\,s)}\right \|_{H^{-\frac{1}{2},-\frac{1}{4}}((\partial D)_T)} - \frac{C\varepsilon}{\| A \|},
\end{eqnarray}
where $C$ is a positive constant.

\medskip
We next prove
\begin{equation}\label{blowup1}
\left \| \partial_\nu \Gamma^0_{(y,\,s)} -\lambda \Gamma^0_{(y,\,s)}\right \|_{H^{-\frac{1}{2},-\frac{1}{4}}((\partial D)_T)}\to \infty\quad \textrm{ as } y\to \partial D.
\end{equation}
Since $\Gamma^0(x,\,t;\,y,\,s) - \Gamma(x,\,t;\,y,\,s)\in C^\infty(\overline \Omega_T)$, that is, $\Gamma^0$ has the same singularity as $\Gamma$ at $(x,\,t)=(y,\,s)$, we only need to prove that
\begin{equation}\label{blowup1-2}
\left \| \partial_\nu \Gamma_{(y,\,s)} -\lambda \Gamma_{(y,\,s)}\right \|_{H^{-\frac{1}{2},-\frac{1}{4}}((\partial D)_T)}\to \infty\quad \textrm{ as } y\to \partial D.
\end{equation}

We consider here the case for $n=3$. Under the assumption that $\partial D$ is of class $ C^2$, for any point $x_0\in\partial D$ there exists a $C^2$-function $\Phi$ such that $$D\cap B(x_0,\, r)=\left\{ x\in B(x_0,\, r):\,x_3> \Phi(x_1,\, x_2) \right\},$$
where $B(x_0,\, r)$ is the ball with radius $r>0$ and center at $x_0$. We choose a new orthonormal basis $\{e_j\},\,j=1,\,2,\,3,$ centered at $x_0$ with $e_3=-\nu$, where $\nu$ is the unit outward normal vector to the boundary at $x_0$. The vectors $e_1$ and $e_2$ lie in the tangent plane to $\partial D$ at $x_0$. Let $x$ be the local coordinate defined by the basis $\{e_j\}$. We introduce the local transformation of coordinate $\eta=F(x)$ as follows:
\begin{equation*}
\eta^\prime=x^\prime,\,\eta_3=x_3-\Phi(x^\prime) \quad \textrm{ with } x^\prime=(x_1,\,x_2),\,\eta^\prime=(\eta_1,\,\eta_2).
\end{equation*}
We note that $\Phi(0)=\nabla_{x^\prime}\Phi(0)=0$. For $x\in\partial D$ and $y\in D$, let $\eta:=F(x),\,\xi:=F(y)$ with $\eta=(\eta^\prime,\,0)$, $\xi=(0,\,\xi_3)$. Then it holds that
\begin{equation*}
|x-y|^2=|F^{-1}(\eta)-F^{-1}(\xi)|^2\leq c_2 |\eta - \xi|^2=c_2(\xi_3^2 + |\eta^\prime|^2).
\end{equation*}

We now pick up the dominant part of \eqref{blowup1-2}, that is,
\begin{equation}\label{domi}
\Vert\partial_{\eta_3}\Gamma_{(\xi,s)}-\lambda \Gamma_{(\xi,s)}\Vert_{H^{-\frac{1}{2},-\frac{1}{4}}((D_2)_T)},
\end{equation}
and ignore all the other terms which are bounded as $y\to \partial D$, where $D_2:=[-l,\,l]\times[-l,\,l]$ with $0< l \ll 1$. To estimate \eqref{domi}, without loss of generality, we assume $s=0$ and introduce an auxiliary function
\begin{equation*}
\varphi(\eta^\prime,\,t)=c\chi(\eta^\prime,\,t)t^{-\alpha}e^{-c_2\frac{|\eta^\prime |^2}{4t}}
\end{equation*}
with $0<\alpha<1/2$, where $\chi(\eta^\prime,\,t)$ is a smooth cut-off function such that it vanishes near $t=T$ and $\eta_j=\pm l \, (j=1, \,2)$ but equals one in a neighborhood of the origin $(\eta^\prime,\,t)=(0,\,0)$. Let us first show that $\varphi\in H^{1,\frac{1}{4}}((D_2)_T)\subset H^{\frac{1}{2},\frac{1}{4}}((D_2)_T)$ and $\|\varphi\|_{H^{\frac{1}{2},\frac{1}{4}}((D_2)_T)}\leq 1$ for sufficiently small constant $c$.

Indeed, by direct calculations, we have
\begin{eqnarray*}
\varphi_t&=&c\chi(\eta^\prime,\,t)e^{-c_2\frac{|\eta^\prime |^2}{4t}}\left(-\alpha t^{-\alpha-1}+ c_2 \frac{|\eta^\prime|^2}{4t} t^{-\alpha -1}\right) + c\chi_t(\eta^\prime,\,t)t^{-\alpha}e^{-c_2\frac{|\eta^\prime |^2}{4t}},\\
\varphi_{\eta_j}&=&-\frac{1}{2}c c_2 \chi(\eta^\prime,\,t) t^{-\alpha-1} e^{-c_2\frac{|\eta^\prime |^2}{4t}}\eta_j + c\chi_{\eta_j}(\eta^\prime,\,t)t^{-\alpha}e^{-c_2\frac{|\eta^\prime |^2}{4t}}.
\end{eqnarray*}
So, for $0<t\ll1$, $\varphi_t$ and $\varphi_{\eta_j}$ can be estimated by
\begin{equation*}
|\varphi_t|\leq c_3 t^{-\alpha -1} e^{-\tilde c_2\frac{|\eta^\prime |^2}{4t}},\quad |\varphi_{\eta_j}| \leq c_3 t^{-\alpha -1/2} e^{-\tilde c_2\frac{|\eta^\prime |^2}{4t}}.
\end{equation*}
Note that
\begin{equation*}
\int_{\mathbb R^2} \left( e^{-\tilde c_2\frac{|\eta^\prime |^2}{4t}} \right)^2 d\eta^\prime =\int_{\mathbb R^2} e^{-\tilde c_2\frac{|\eta^\prime |^2}{2t}} d\eta^\prime= O(t).
\end{equation*}
Then we have
\begin{equation}\label{bp1}
\int_0^T\int_{D_2} |\varphi_{\eta_j}|^2dx dt \leq c_4 \int_0^T (t^{-\alpha-1/2})^2\, t\, dt = c_4\int_0^T t^{-2\alpha}\, dt <\infty.
\end{equation}
In addition, we deduce that
\begin{eqnarray}\label{bp2}
\left(\int_{D_2} [\partial_t^{1/4} \varphi]^2 dx\right)^{1/2} & \leq & \left\{\int_{\mathbb R^2} \left( \frac{1}{\Gamma(3/4)}\int_0^t (t-\tau)^{-1/4}\partial_\tau\varphi(x,\,\tau)d\tau\right)^2 dx \right\}^{1/2}\nonumber\\
&\leq& \frac{1}{\Gamma(3/4)} \int_0^t (t-\tau)^{-1/4} \left(\int_{\mathbb R^2} |\partial_\tau\varphi(x,\,\tau)|^2 dx \right)^{1/2}d\tau\nonumber\\
&\leq& c_5\int_0^t (t-\tau)^{-1/4} \tau^{-\alpha-1/2} d\tau\nonumber\\
&\leq& c_6 t^{-\alpha+1/4},
\end{eqnarray}
which guarantees the integrability of $\partial_t^{1/4} \varphi$ in $L^2$ under our condition $0<\alpha<1/2$. Thus, we conclude that $\varphi\in H^{1,\frac{1}{4}}((D_2)_T)\subset H^{\frac{1}{2},\frac{1}{4}}((D_2)_T)$ and $\|\varphi\|_{H^{\frac{1}{2},\frac{1}{4}}((D_2)_T)}\leq 1$ by taking a sufficiently small constant $c$.

Next, we compute the norm of $\partial_{\eta_3}\Gamma_{(\xi,s)} -\lambda \Gamma_{(\xi,s)}$ using the duality. In fact, we have
\begin{eqnarray*}
&&\left\Vert\partial_{\eta_3}\Gamma_{(\xi,s)} -\lambda \Gamma_{(\xi,s)} \right\Vert_{H^{-\frac{1}{2},-\frac{1}{4}}((\partial D)_T)}\\
&\ge& \frac{c_7}{8\pi^{3/2}}\int_0^T \int_{D_2} \left(\frac{c_2}{2t}\xi_3-\lambda_2\right) t^{-3/2} e^{-c_2\frac{\xi_3^2+|\eta^\prime|^2}{4t}}\left( c \chi t^{-\alpha}e^{-c_2\frac{|\eta^\prime|^2}{4t}} \right)d\eta^\prime dt\\
&\geq&  \frac{c_8}{8\pi^{3/2}}\int_0^T  \left( \frac{c_2}{2t}\xi_3 -\lambda_2 \right) t^{-1/2-\alpha} e^{-c_2\frac{\xi^2_3}{4t}}dt\\
&=& \frac{c_8}{8\pi^{3/2}} \int^\infty_{\frac{\xi_3^2}{T}} \left( \frac{c_2\tau}{2}\xi_3^{-2\alpha}  -\lambda_2\xi_3^{-2\alpha+1}  \right)  \tau^{\alpha -\frac{3}{2}} e^{-c_2\frac{\tau}{4}} d\tau \to \infty\quad  \textrm{ as }  \xi_3\to 0\; (y\to\partial D).
\end{eqnarray*}

Now we can conclude from \eqref{badd1} and \eqref{blowup1} that the blow-up properties \eqref{eq:density11} and \eqref{eq:density12} hold. The proof is complete.  \hfill $\Box$

\bigskip
To further investigate the behavior of the density $g^y$ as $y$ approaches to $\partial D$ from the exterior of $D$, we need the following lemma.

\begin{lemma}\label{lem:5}
For any fixed $s\in(0,\, T)$, the Green function $\Gamma^0_{(y,\, s)}(x,\,t)$ is not in the range of $A$ if $y\in \Omega\setminus D$.
\end{lemma}
{\bf Proof.} Suppose that $\Gamma^0_{(y,\, s)}(x,\,t)$ is in the range of $A$. Then there exists a function $g\in H^{-\frac{1}{2},-\frac{1}{4}}((\partial D)_T)$ such that
\begin{equation}\label{ad1}
w^g|_{(\partial\Omega)_T}=\Gamma^0_{(y,\,s)}|_{(\partial\Omega)_T},
\end{equation}
where $w^g\in \tilde H^{1,\frac{1}{2}}((\Omega\setminus\overline D)_T)$ is the solution to
\begin{equation*}
\left\{ \begin{array}{l}
(\partial_t - \Delta) w^g=0 \quad \textrm{ in }(\Omega\setminus\overline D)_T,\\
\partial_\nu w^g -\lambda w^g =g \quad \textrm{ on }(\partial D)_T,\\
\partial_\nu w^g=0 \quad \textrm{ on }(\partial\Omega)_T,\\
w^g=0 \quad \textrm{ at } t=0.
\end{array}\right.
\end{equation*}
From \eqref{ad1} and $\partial_\nu w^g|_{(\partial\Omega)_T}=\partial_\nu \Gamma^0_{(y,\,s)}|_{(\partial\Omega)_T}=0$, it follows that $w^g=\Gamma^0_{(y,\,s)}$ in $\left(\Omega\setminus(\overline D \cup \{y\})\right)_T$
and hence $\| \Gamma^0_{(y,\,s)} \|_{ \tilde H^{1,\frac{1}{2}}((\Omega\setminus\overline D)_T) } =\| w^g \|_{\tilde H^{1,\frac{1}{2}}((\Omega\setminus\overline D)_T)}<\infty$. Here $y\in \Omega\setminus D$ can be either $y\in \partial D$ or $y\in\Omega\setminus\overline D$, and for each case it holds $\| \Gamma^0_{(y,\, s)} \|_{ \tilde H^{1,\frac{1}{2}}((\Omega\setminus\overline D)_T) } =\infty$. This gives a contradiction and completes the proof. \hfill $\Box$

\bigskip
In contrast to Theorem \ref{prop:3} for $y\in D$, we now establish the following blowup property of the density $g^y$ for $y\not\in D$.
\begin{theorem}\label{prop:6}
Fix $s\in (0,\, T)$  and let $y\in \Omega\setminus D$. Then, for every $\varepsilon > 0$ and $\delta>0$, there exists a function $g^y_{\varepsilon,\,\delta}\in H^{-\frac{1}{2},-\frac{1}{4}}((\partial\Omega)_T)$ satisfying
\begin{equation}\label{eq:data2}
\| Fg^y_{\varepsilon,\,\delta} - \Gamma^0_{(y,\,s)} \|_{H^{\frac{1}{2},\frac{1}{4}}((\partial\Omega)_T)}< \varepsilon+\delta
\end{equation}
such that
\begin{equation}\label{eq:blowup1}
\lim_{\delta\to 0} \| g^y_{\varepsilon,\,\delta} \|_{H^{-\frac{1}{2},-\frac{1}{4}}((\partial\Omega)_T) }=\infty
\end{equation}
and
\begin{equation}\label{eq:blowup2}
\lim_{\delta\to 0} \| S g^y_{\varepsilon,\,\delta} \|_{\tilde H^{1,\frac{1}{2}}(D_T)}=\infty.
\end{equation}
\end{theorem}
{\bf Proof.} By Lemma \ref{lem:4}, for arbitrary $\delta>0$ there exists a function $f^y_\delta\in H^{-\frac{1}{2},-\frac{1}{4}}((\partial D)_T)$
such that
\begin{equation*}
\| A f^y_\delta - \Gamma^0_{(y,\, s)} \|_{ H^{\frac{1}{2},\frac{1}{4}}((\partial\Omega)_T) }< \delta.
\end{equation*}
Since $\Gamma^0_{(y,\, s)}$ is not in the range of $A$ due to Lemma \ref{lem:5}, we have
 \begin{equation}\label{add32}
\| f^y_\delta \|_{ H^{-\frac{1}{2},-\frac{1}{4}}((\partial D)_T)}\to\infty \quad \textrm{ as } \delta\to 0.
\end{equation}
Recall that the range of $H$ is dense in $ H^{-\frac{1}{2},-\frac{1}{4}}((\partial D)_T)$. We can find a function $g^y_{\varepsilon,\,\delta}\in H^{-\frac{1}{2},-\frac{1}{4}}((\partial\Omega)_T)$ such that
\begin{equation}\label{add3}
\|H g^y_{\varepsilon,\,\delta} + f^y_\delta \|_{H^{-\frac{1}{2},-\frac{1}{4}}((\partial D)_T)}< \varepsilon/(\|A\|+1),
\end{equation}
and hence
\begin{equation*}
\| A f^y_\delta + A H g^y_{\varepsilon,\,\delta} \|_{ H^{\frac{1}{2},\frac{1}{4}}((\partial\Omega)_T)}< \varepsilon.
\end{equation*}
So, it can be derived that
\begin{eqnarray*}
\| F g^y_{\varepsilon,\,\delta} - \Gamma^0_{(y,\, s)}\|_{H^{\frac{1}{2},\frac{1}{4}}((\partial\Omega)_T)} &=& \| -AH g^y_{\varepsilon,\,\delta} - \Gamma^0_{(y,\, s)}\|_{ H^{\frac{1}{2},\frac{1}{4}}((\partial\Omega)_T) }\\
&\leq & \| AH g^y_{\varepsilon,\,\delta} + Af^y_\delta \|_{ H^{\frac{1}{2},\frac{1}{4}}((\partial\Omega)_T) } + \| Af^y_\delta - \Gamma^0_{(y,\, s)}\|_{ H^{\frac{1}{2},\frac{1}{4}}((\partial\Omega)_T) }\\
&\leq & \varepsilon+\delta.
\end{eqnarray*}
Moreover, we obtain from \eqref{add32} and \eqref{add3} that
\begin{equation*}
\| Hg^y_{\varepsilon,\,\delta} \|_{H^{-\frac{1}{2},-\frac{1}{4}}((\partial D)_T)}\to \infty \quad \textrm{ as } \delta\to 0,
\end{equation*}
which implies
\begin{equation*}
\| g^y_{\varepsilon,\,\delta} \|_{H^{-\frac{1}{2},-\frac{1}{4}}((\partial\Omega)_T)}\to \infty,\quad \|S g^y_{\varepsilon,\,\delta} \|_{\tilde H^{1,\frac{1}{2}}(D_T)}\to \infty \quad \textrm{ as } \delta\to 0.
\end{equation*}
This completes the proof. \hfill $\Box$

\bigskip
By Theorems \ref{prop:3} and \ref{prop:6}, $Sg^y$ can be taken as an indicator function for reconstructing the boundary of the cavity $D$, where $g^y$ is the approximate solution to \eqref{gap} for the sampling point $y\in\Omega$.

\section{The asymptotic behavior of the indicator function}\label{asy}
\setcounter{equation}{0}

In this section, we show a short time asymptotic behavior of the indicator function $Sg^y$ as $y$ approaches to the boundary $\partial D$ from the interior of $D$. To do this, we analyze the asymptotic behavior of the reflected solution of the fundamental solution. As we know, the reflected solution is the compensating term of the Green function for the related initial-boundary value problem. This means that we actually provide a pointwise short time asymptotic behavior of the Green function near $\partial D$. From our asymptotic analysis, we can establish a pointwise reconstruction scheme for the location and shape of the cavity, and can also know the distance to the unknown cavity when we probe it from the interior of $D$. Without loss of generality, we only consider the three dimensional case.

Let $\bar \Gamma_{(y,\,s)}(x,\,t)$ be the reflected solution of $\Gamma^0_{(y,\,s)}(x,\,t)$. That is,
\begin{equation*}
\left\{ \begin{array}{l}
(\partial_t - \Delta) \bar \Gamma_{(y,\,s)}(x,\,t)=0 \quad \textrm{ in } D_T,\\
(\partial_\nu -\lambda) \bar \Gamma_{(y,\,s)}(x,\,t) =-(\partial_\nu -\lambda)\Gamma^0_{(y,\,s)}(x,\,t)\quad \textrm{ on }(\partial D)_T,\\
\bar \Gamma_{(y,\,s)}(x,\,t)=0 \quad \textrm{ for } x\in D,\, t\leq s.
\end{array}\right.
\end{equation*}
Then, by the well-posedness of the above initial-boundary value problem, we obtain from \eqref{badd0} that
\begin{equation}\label{asi1}
\left\|Sg^y - \bar \Gamma_{(y,\,s)}\right\|_{\tilde H^{1,\frac{1}{2}}(D_T)} < C_1\varepsilon,
\end{equation}
where $C_1$ is a positive constant. Denote by $R_{(y,\,s)}(x,\,t)$ the reflected solution of $\Gamma_{(y,s)}(x,\,t)$, i.e.,
\begin{equation}\label{si1}
\left\{ \begin{array}{l}
(\partial_t - \Delta) R_{(y,\,s)}(x,\,t)=0 \quad \textrm{ in } D_T,\\
(\partial_\nu -\lambda) R_{(y,\,s)}(x,\,t) =-(\partial_\nu -\lambda)\Gamma_{(y,\,s)}(x,\,t)\quad \textrm{ on }(\partial D)_T,\\
R_{(y,\,s)}(x,\,t)=0 \quad \textrm{ for } x\in D,\, t\leq s.
\end{array}\right.
\end{equation}
Since $\Gamma^0_{(y,\,s)}(x,\,t)-\Gamma_{(y,\,s)}(x,\,t)$ for $(y,\,s)\in \overline D_T$ is smooth in $\Omega_T$, there exists a positive constant $C_2$ such that
\begin{equation}\label{asi2}
\left\|\bar \Gamma_{(y,\,s)}(x,\,t) - R_{(y,\,s)}(x,\,t) \right\|_{\tilde H^{1,\frac{1}{2}}(D_T)} < C_2.
\end{equation}
Combining \eqref{asi1} and \eqref{asi2} yields
\begin{equation}\label{asi12}
\left\|\left(Sg^y\right)(x,\,t) - R_{(y,\,s)}(x,\,t) \right\|_{\tilde H^{1,\frac{1}{2}}(D_T)} < C_1\varepsilon + C_2.
\end{equation}
This implies that $Sg^y$ and $R_{(y,\,s)}$ have the same blow-up behavior as $y$ approaches to $\partial D$. Hence, to get the asymptotic behavior of $Sg^y$, we only need to show the corresponding result of $R_{(y,\,s)}(x,\,t)$.

\medskip
To begin with, we locally flatten the boundary $\partial D$. For any point $z\in\partial D$, there is a diffeomorphism $\Phi:\,\mathbb R^3 \to \mathbb R^3$ which transforms $z$ to the origin ${\bf 0}$ such that $$\Phi(D)\subset \mathbb R_-^3:=\left\{\xi=(\xi_1,\,\xi_2,\,\xi_3)\in\mathbb R^3: \; \xi_3 <0 \right\}.$$ Without loss of generality, we assume ${\bf 0}\in\partial D$ with $\nu({\bf 0})=e_3$ and locally flatten the boundary $\partial D$ around ${\bf 0}$. Let $\xi=\Phi(x)$ and $\eta=\Phi(y)$. Denote by $J(x)$ the Jacobian matrix of $\Phi$. We have
$$J(x)=\nabla \Phi(x),\; J({\bf 0})=I,\; \nu(x) =\frac{J^T\nu(\xi)}{ |J^T\nu(\xi)|}.$$
Set $\tilde R(\xi,\,t;\,\eta,\,s):=R(x,\,t;\,y,\,s)$ and $\tilde \Gamma(\xi,\,t;\,\eta,\,s):=\Gamma(x,\,t;\,y,\,s)$. We deduce from \eqref{si1} that $\tilde R(\xi,\,t;\,\eta,\,s)$ near ${\bf 0}$ satisfies
\begin{equation}\label{si33}
\left\{ \begin{array}{l}
(\partial_t - \nabla_\xi \cdot M(\xi)\nabla_\xi) \tilde R(\xi,\,t;\,\eta,\,s)=0, \quad t\in (0,\,T),\;\xi_3<0,\\
e_3\cdot(M(\xi)\nabla_\xi \tilde R)(\xi,\,t;\,\eta,\,s)-\lambda(\Phi^{-1}(\xi)) |J^T\nu(\xi)|\tilde R(\xi,\,t;\,\eta,\,s)\\
\quad\quad=-\left(e_3\cdot(M(\xi)\nabla_\xi \tilde \Gamma)(\xi,\,t;\,\eta,\,s)-\lambda(\Phi^{-1}(\xi))|J^T\nu(\xi)| \tilde \Gamma(\xi,\,t;\,\eta,\,s)\right)
,\quad t\in (0,\,T),\;\xi_3=0,\\
\tilde R(\xi,\,t;\,\eta,\,s)=0, \quad t\leq s,\;\xi_3<0,
\end{array}\right.
\end{equation}
where $M(\xi)=(JJ^T)(\Phi^{-1}(\xi))$. Let $W^+(\xi,\,t;\,\eta,\,s)$ be such that
\begin{equation}\label{si44}
\left\{ \begin{array}{l}
(\partial_t - \Delta_\xi) W^+(\xi,\,t;\,\eta,\,s)=0, \quad t\in (0,\,T),\;\xi_3<0,\\
\left(\partial_{\xi_3}-\lambda({\bf 0})\right) W^+(\xi,\,t;\,\eta,\,s)=-\left(\partial_{\xi_3}-\lambda({\bf 0})\right) \tilde \Gamma(\xi,\,t;\,\eta,\,s),\quad t\in (0,\,T),\;\xi_3=0,\\
W^+(\xi,\,t;\,\eta,\,s)=0, \quad t\leq s,\;\xi_3<0.
\end{array}\right.
\end{equation}

In the following, we first estimate the difference $\tilde R(\xi,\,t;\,\eta,\,s) - W^+(\xi,\,t;\,\eta,\,s)$, and then analyze the asymptotic behavior of $W^+(\xi,\,t;\,\eta,\,s)$. As a consequence, the asymptotic behavior of $\tilde R(\xi,\,t;\,\eta,\,s)$ is obtained.

\begin{lemma}\label{thas0}
Let $\eta=(\eta_1,\,\eta_2,\,\eta_3)^T$, where $\eta_3=-\epsilon/2$ and $|\eta|\leq c\epsilon$ for small $\epsilon>0$ and some constant $c>0$. Then there exists a positive constant $C$ such that
\begin{equation}\label{do1}
\left|\tilde R(\eta,\,s+\epsilon^2;\,\eta,\,s) - W^+(\eta,\,s+\epsilon^2;\,\eta,\,s)\right|\leq C \epsilon^{-2} \quad \textrm{ as } \epsilon\to 0.
\end{equation}
\end{lemma}
{\bf Proof.} Define $P(\xi,\,t;\,\eta,\,s):=\tilde R(\xi,\,t;\,\eta,\,s) - W^+(\xi,\,t;\,\eta,\,s).$ We derive from \eqref{si33} and \eqref{si44} that near ${\bf 0}$ the function $P(\xi,\,t;\,\eta,\,s)$ satisfies
\begin{equation}\label{si41}
\left\{ \begin{array}{l}
(\partial_t - \Delta_\xi) P(\xi,\,t;\,\eta,\,s)=\nabla\cdot (M-I)\nabla \tilde R, \quad t\in (0,\,T),\;\xi_3<0,\\
\left(\partial_{\xi_3}-\lambda({\bf 0})\right) P(\xi,\,t;\,\eta,\,s)=\left(\partial_{\xi_3}(\tilde R+\tilde \Gamma)-e_3\cdot[M(\xi)\nabla_\xi(\tilde R+\tilde\Gamma)]\right)\\
\quad\quad\quad\quad\quad\quad\quad\quad\quad\quad\quad\quad-\left(\lambda({\bf 0})-\lambda(\Phi^{-1}(\xi))|J^T\nu(\xi)|\right)(\tilde R + \tilde \Gamma),\quad t\in (0,\,T),\;\xi_3=0,\\
P(\xi,\,t;\,\eta,\,s)=0, \quad t\leq s,\;\xi_3<0.
\end{array}\right.
\end{equation}
Denote by $\hat\Gamma_{(\eta,\,s)}(\xi,\,t):=\hat\Gamma(\xi,\,t;\,\eta,\,s)$ the solution to the following problem:
\begin{equation}\label{si42}
\left\{ \begin{array}{l}
-(\partial_t + \Delta_\xi) \hat\Gamma(\xi,\,t;\,\eta,\,s)=\delta(t-s)\delta(\xi-\eta), \quad t\in (0,\,T),\;\xi_3<0,\\
\left(\partial_{\xi_3}-\lambda({\bf 0})\right) \hat\Gamma(\xi,\,t;\,\eta,\,s)=0,\quad t\in (0,\,T),\;\xi_3=0,\\
\hat\Gamma(\xi,\,t;\,\eta,\,s)=0, \quad t\geq s,\;\xi_3<0.
\end{array}\right.
\end{equation}
Let $Q$ be a small bounded domain in the lower half-plane such that ${\bf 0}\in\partial Q\cap \partial \mathbb R^3_-\subset\partial (\Phi(D))$ and $Q$ totally lies in $\Phi(D)$. We decompose the boundary of $Q$ into two disjoint parts such that $\partial Q=\partial Q_1\cup \partial Q_2$ with $\partial Q_1=\partial Q\cap \partial \mathbb R^3_-$. Since we only consider the singularity locally at ${\bf 0}$, we have
\begin{eqnarray}\label{do2}
P_{(\eta,\,s)}(\xi,\,t)&=&-\int_s^td\tau \int_{\partial Q} \left( P_{(\eta,\,s)}(z,\,\tau) \partial_{\nu(z)} \hat\Gamma_{(\xi,\,t)}(z,\,\tau)  - \partial_{\nu(z)} P_{(\eta,\,s)}(z,\,\tau)  \hat\Gamma_{(\xi,\,t)}(z,\,\tau)\right)d\sigma(z)\nonumber\\
& & + \int_s^td\tau \int_{\partial Q} \left( \nu(z)\cdot(M(z)-I)\nabla\tilde R_{(\eta,\,s)}(z,\,\tau)\cdot \hat\Gamma_{(\xi,\,t)}(z,\,\tau) \right)ds(z)\nonumber\\
& &-\int_s^td\tau \int_Q \left( (M(z)-I)\nabla\tilde R_{(\eta,\,s)}(z,\,\tau)\cdot \nabla\hat\Gamma_{(\xi,\,t)}(z,\,\tau) \right)dz \nonumber\\
&=&-\int_s^td\tau \int_{\partial Q_1} \left( P_{(\eta,\,s)}(z,\,\tau) \partial_{z_3} \hat\Gamma_{(\xi,\,t)}(z,\,\tau)  - \partial_{z_3} P_{(\eta,\,s)}(z,\,\tau)  \hat\Gamma_{(\xi,\,t)}(z,\,\tau)\right)dz^\prime\nonumber\\
& & + \int_s^td\tau \int_{\partial Q_1} \left( e_3\cdot(M(z)-I)\nabla\tilde R_{(\eta,\,s)}(z,\,\tau)\cdot \hat\Gamma_{(\xi,\,t)}(z,\,\tau) \right)ds(z)\nonumber\\
& &-\int_s^td\tau \int_Q \left( (M(z)-I)\nabla\tilde R_{(\eta,\,s)}(z,\,\tau)\cdot \nabla\hat\Gamma_{(\xi,\,t)}(z,\,\tau) \right)dz + O(1).
\end{eqnarray}
To estimate \eqref{do2}, we need the following estimates \cite{Fri1964, Gre1971}:
\begin{equation*}
\begin{array}{ll}
\left| \tilde R_{(\eta,\,s)}(z,\,\tau) \right| \leq  c_1(\tau - s)^{-3/2}\exp\left( -\frac{|z-\hat\eta|^2}{c_2(\tau -s )} \right), & \left|\nabla_z \tilde R_{(\eta,\,s)}(z,\,\tau) \right| \leq  c_3(\tau - s)^{-2}\exp\left( -\frac{|z-\hat\eta|^2}{c_4(\tau -s )} \right),\\
\left| \tilde \Gamma_{(\eta,\,s)}(z,\,\tau) \right| \leq  c_1(\tau - s)^{-3/2}\exp\left( -\frac{|z-\eta|^2}{c_2(\tau -s )} \right),&\left| \nabla_z\tilde \Gamma_{(\eta,\,s)}(z,\,\tau) \right| \leq  c_3(\tau - s)^{-2}\exp\left( -\frac{|z-\eta|^2}{c_4(\tau -s )} \right),\\
\left| \hat\Gamma_{(\xi,\,t)}(z,\,\tau) \right| \leq  c_1(t-\tau)^{-3/2}\exp\left( -\frac{|z-\xi|^2}{c_2(t-\tau )} \right),& \left|\nabla_z \hat\Gamma_{(\xi,\,t)}(z,\,\tau) \right| \leq  c_3(t-\tau)^{-2}\exp\left( -\frac{|z-\xi|^2}{c_4(t-\tau )} \right),
\end{array}
\end{equation*}
where $\hat\eta=(\eta_1,\,\eta_2,\,-\eta_3)^T$ and $c_j\,(j=1,\,\cdots,\,4)$ are positive constants.

On the plane $z_3=0$, it holds that
\begin{eqnarray}\label{do22}
& &\left|P_{(\eta,\,s)}(z,\,\tau) \partial_{z_3} \hat\Gamma_{(\xi,\,t)}(z,\,\tau)  - \partial_{z_3} P_{(\eta,\,s)}(z,\,\tau)  \hat\Gamma_{(\xi,\,t)}(z,\,\tau)\right|\nonumber\\
&=&\left|\left(\partial_{z_3}(\tilde R+\tilde\Gamma)-e_3\cdot[M(z)\nabla_z(\tilde R+\tilde\Gamma)]\right)\hat\Gamma_{(\xi,\,t)}(z,\,\tau)\right|\nonumber\\
&&+\left|\left(\lambda({\bf 0})-\lambda(\Phi^{-1}(z))|J^T\nu(z)|\right)(\tilde R+\tilde \Gamma)\,\hat\Gamma_{(\xi,\,t)}(z,\,\tau)\right|\nonumber\\
&=&\left|\left(e_3\cdot[(I-M(z))\nabla_z(\tilde R+\tilde\Gamma)]\right)\hat\Gamma_{(\xi,\,t)}(z,\,\tau)\right|\nonumber\\
&&+\left|\left[\lambda({\bf 0})(1-|J^T\nu(z)|)+(\lambda({\bf 0})-\lambda(\Phi^{-1}(z)))|J^T\nu(z)|\right](\tilde R+\tilde \Gamma)\,\hat\Gamma_{(\xi,\,t)}(z,\,\tau)\right|\nonumber\\
&\leq& C_0|z^\prime|\left(\left|\nabla\tilde R_{(\eta,\,s)}(z,\,\tau)\right|+\left|\nabla\tilde \Gamma_{(\eta,\,s)}(z,\,\tau)\right|+\left|\tilde R_{(\eta,\,s)}(z,\,\tau)\right|+\left|\tilde \Gamma_{(\eta,\,s)}(z,\,\tau)\right|\right)\,\left|\hat\Gamma_{(\xi,\,t)}(z,\,\tau)\right|.
\end{eqnarray}
Set $z^\prime:=(z_1,\,z_2)$, $\eta^\prime:=(\eta_1,\,\eta_2)$ and $\xi^\prime:=(\xi_1,\,\xi_2)$. Using Lemma 3 in \cite[Chapter 1]{Fri1964}, we derive that
\begin{eqnarray}\label{do3}
&&\int_s^td\tau \int_{\partial Q_1} |z^\prime|\left(\left|\nabla\tilde R_{(\eta,\,s)}(z,\,\tau)\right|+\left|\nabla\tilde \Gamma_{(\eta,\,s)}(z,\,\tau)\right|\right)\, \left|\hat\Gamma_{(\xi,\,t)}(z,\,\tau)\right|dz^\prime\nonumber\\
&\leq&C_1\int_s^td\tau \int_{\partial Q_1}\left( |z^\prime-\eta^\prime| + |\eta^\prime| \right)\,\left(\left|\nabla\tilde R_{(\eta,\,s)}(z,\,\tau)\right|+\left|\nabla\tilde \Gamma_{(\eta,\,s)}(z,\,\tau)\right|\right)\, \left|\hat\Gamma_{(\xi,\,t)}(z,\,\tau)\right|dz^\prime\nonumber\\
&\leq&C_2\int_s^td\tau \int_{\partial Q_1} |z^\prime-\eta^\prime|\,(\tau - s)^{-2}\exp\left( -\frac{|z^\prime-\eta^\prime|^2+\eta_3^2}{c_2(\tau -s )}\right)\, (t- \tau )^{-3/2}\exp\left( -\frac{|z^\prime-\xi^\prime|^2+\xi_3^2}{c_2(t- \tau)}\right)dz^\prime\nonumber\\
&& + C_3\int_s^td\tau \int_{\partial Q_1} |\eta^\prime|\,(\tau - s)^{-2}\exp\left( -\frac{|z^\prime-\eta^\prime|^2+\eta_3^2}{c_2(\tau -s )}\right)\, (t-\tau)^{-3/2}\exp\left( -\frac{|z^\prime-\xi^\prime|^2+\xi_3^2}{c_2(t- \tau)}\right)dz^\prime\nonumber\\
&\leq&C_4\eta_3^{-2}\xi_3^{-2}\int_s^td\tau \int_{\partial Q_1} (\tau - s)^{-1/2}\exp\left( -\frac{|z^\prime-\eta^\prime|^2}{c_7(\tau -s )}\right)\, (t- \tau )^{-1/2}\exp\left( -\frac{|z^\prime-\xi^\prime|^2}{c_7(t- \tau)}\right)dz^\prime\nonumber\\
&& + C_5|\eta^\prime|\eta_3^{-2}\xi_3^{-2}\int_s^td\tau \int_{\partial Q_1} (\tau - s)^{-1}\exp\left( -\frac{|z^\prime-\eta^\prime|^2}{c_2(\tau -s )}\right)\, (t-\tau)^{-1/2}\exp\left( -\frac{|z^\prime-\xi^\prime|^2}{c_2(t- \tau)}\right)dz^\prime\nonumber\\
&\leq&C_6\eta_3^{-2}\xi_3^{-2}(t-s)\exp\left( -\frac{|\xi^\prime-\eta^\prime|^2}{c_7(t-s)} \right)+C_7|\eta^\prime|\eta_3^{-2}\xi_3^{-2}(t-s)^{1/2}\exp\left( -\frac{|\xi^\prime-\eta^\prime|^2}{c_2(t-s)} \right).
\end{eqnarray}
Noticing that
\begin{equation*}
\int_s^td\tau \int_{\partial Q_1} |z^\prime|\left(\left|\tilde R_{(\eta,\,s)}(z,\,\tau)\right|+\left|\tilde \Gamma_{(\eta,\,s)}(z,\,\tau)\right|\right)\, \left|\hat\Gamma_{(\xi,\,t)}(z,\,\tau)\right|dz^\prime.
\end{equation*}
has a less singularity than \eqref{do3}, we omit the estimate of this integral here.

In addition, the integral
\begin{equation*}
\int_s^td\tau \int_{\partial Q_1} \left( e_3\cdot(M(z)-I)\nabla\tilde R_{(\eta,\,s)}(z,\,\tau)\cdot \hat\Gamma_{(\xi,\,t)}(z,\,\tau) \right)ds(z)
\end{equation*}
can be estimated in the same way as in \eqref{do3}. As for the volume integral in \eqref{do2}, we have that
\begin{eqnarray}\label{do4}
&&\left|\int_s^td\tau \int_Q \left( (M(z)-I)\nabla\tilde R_{(\eta,\,s)}(z,\,\tau)\cdot \nabla\hat\Gamma_{(\xi,\,t)}(z,\,\tau) \right)dz\right|\nonumber\\
&\leq&C_8\int_s^td\tau \int_Q |z|\left|\nabla\tilde R_{(\eta,\,s)}(z,\,\tau)\right|\, \left|\nabla\hat\Gamma_{(\xi,\,t)}(z,\,\tau) \right|dz\nonumber\\
&\leq&C_9(t-s)^{-1}\exp\left( -\frac{|\xi-\hat\eta|^2}{c_4(t-s)} \right)+C_{10}|\eta|(t-s)^{-3/2}\exp\left( -\frac{|\xi-\hat\eta|^2}{c_4(t-s)} \right)\nonumber\\
&\leq&C_{9}(t-s)^{-1}\exp\left( -\frac{|\xi_3+\eta_3|^2}{c_4(t-s)} \right)+C_{10}|\eta|(t-s)^{-3/2}\exp\left( -\frac{|\xi_3+\eta_3|^2}{c_4(t-s)} \right).
\end{eqnarray}
Take $\xi=\eta$ with $|\eta|\leq c\epsilon$, $\xi_3+\eta_3=-\epsilon,\;t-s=\epsilon^2$.
Then we finally get the following estimates for $\epsilon\to 0$:
\begin{eqnarray*}
&&\left|\int_s^td\tau \int_{\partial Q_1} \left( P_{(\eta,\,s)}(z,\,\tau) \partial_3 \hat\Gamma_{(\xi,\,t)}(z,\,\tau)  - \partial_3 P_{(\eta,\,s)}(z,\,\tau)  \hat\Gamma_{(\xi,\,t)}(z,\,\tau)\right)dz^\prime\right|\leq C_{11}\epsilon^{-2}, \\
&&\left|\int_s^td\tau \int_Q \left( (M(z)-I)\nabla\tilde R_{(\eta,\,s)}(z,\,\tau)\cdot \nabla\hat\Gamma_{(\xi,\,t)}(z,\,\tau) \right)dz\right|\leq C_{12}\epsilon^{-2},
\end{eqnarray*}
which lead to the estimate \eqref{do1} by \eqref{do2}. The proof of this lemma is complete. \hfill $\Box$

\bigskip
Next we derive the expression of $W^+(\xi,\,t;\,\eta,\,s)$, and then show its asymptotic behavior.

\begin{lemma}\label{thas1}
The solution $W^+(\xi,\,t;\,\eta,\,s)$ to \eqref{si44} can be expressed by
\begin{equation}\label{is16}
W^+(\xi,\,t;\,\eta,\,s)=\frac{1}{16\pi^3i} \int_{\sigma - i \infty}^{\sigma+i\infty} e^{t\tau} d\tau \int_{\mathbb R^2} e^{i\xi^\prime\cdot \zeta^\prime} \frac{\Theta +\lambda_0}{\Theta (\Theta - \lambda_0)} e^{-(\tau s + i \zeta^\prime\cdot\eta^\prime)} e^{\Theta (\xi_3+\eta_3)}d\zeta^\prime,
\end{equation}
where $\xi^\prime:=(\xi_1,\,\xi_2),\,\eta^\prime:=(\eta_1,\,\eta_2)$, $\lambda_0:=\lambda({\bf 0})$ and $\Theta:=\sqrt{\tau+|\zeta^\prime|^2}$ with $\mathrm{Re}\,\Theta\geq0$.
\end{lemma}
{\bf Proof.} Define $H(\xi,\,t;\,\eta,\,s):=W^+(\xi,\,t;\,\eta,\,s)+\tilde\Gamma(\xi,\,t;\,\eta,\,s)$. We know from \eqref{si44} that $H(\xi,\,t;\,\eta,\,s)$ satisfies
\begin{equation}\label{si4}
\left\{ \begin{array}{l}
(\partial_t - \Delta_\xi) H(\xi,\,t;\,\eta,\,s)=\delta(t-s)\delta(\xi-\eta), \quad t\in (0,\,T),\;\xi_3<0,\\
\left(\partial_{\xi_3}-\lambda({\bf 0})\right) H(\xi,\,t;\,\eta,\,s)=0,\quad t\in (0,\,T),\;\xi_3=0,\\
H(\xi,\,t;\,\eta,\,s)=0, \quad t\leq s,\;\xi_3<0.
\end{array}\right.
\end{equation}
Denote by $\hat H$ the Laplace transform of $H$ with respect to $t$. Then we have
\begin{equation}\label{si5}
\left\{ \begin{array}{l}
(\tau - \Delta_\xi) \hat H(\xi,\,t;\,\eta,\,s)=e^{-\tau s}\delta(\xi-\eta), \quad \xi_3<0,\\
(\partial_{\xi_3}-\lambda_0) \hat H(\xi,\,t;\,\eta,\,s)=0,\quad\xi_3=0.\\
\end{array}\right.
\end{equation}
We now look for the solution to \eqref{si5} in the form of $$\hat H(\xi,\,t;\,\eta,\,s)=\hat H^{\pm}(\xi,\,t;\,\eta,\,s),\quad \pm(\xi_3 - \eta_3)>0,$$ where $\hat H^{\pm}(\xi,\,t;\,\eta,\,s)$ satisfy
\begin{equation}\label{si6}
\left\{ \begin{array}{l}
(\tau - \Delta_\xi) \hat H^{\pm}(\xi,\,t;\,\eta,\,s)=0, \quad \pm(\xi_3 - \eta_3)>0,\\
\hat H^+ - \hat H^-=0, \quad \xi_3=\eta_3,\\
\partial_{\xi_3} (\hat H^+ - \hat H^-)=-e^{-\tau s} \delta(\xi^\prime - \eta^\prime), \quad \xi_3=\eta_3,\\
(\partial_{\xi_3}-\lambda_0) \hat H^+(\xi,\,t;\,\eta,\,s)=0,\quad\xi_3=0.\\
\end{array}\right.
\end{equation}
Denote by $\varphi^\pm$ the Fourier transform of $\hat H^\pm$ with respect to $\xi^\prime=(\xi_1,\,\xi_2)$ and let $\zeta^\prime=(\zeta_1,\,\zeta_2)$ be the Fourier variable corresponding to $\xi^\prime$. Then we have
\begin{equation}\label{si7}
\left\{ \begin{array}{l}
(\tau + |\zeta^\prime|^2 - \partial^2_{\xi_3}) \varphi^\pm=0, \quad \pm(\xi_3 - \eta_3)>0,\\
\varphi^+ - \varphi^-=0, \quad \xi_3=\eta_3,\\
\partial_{\xi_3} (\varphi^+ - \varphi^-)=-e^{-\tau s} e^{-i\zeta^\prime\cdot\eta^\prime}, \quad \xi_3=\eta_3,\\
(\partial_{\xi_3}-\lambda_0) \varphi^+=0,\quad\xi_3=0.\\
\end{array}\right.
\end{equation}
Since the operator $\partial^2_{\xi_3}-\Theta^2$ has the fundamental solutions $e^{\pm \Theta \xi_3}$, we seek the solution to \eqref{si7} in the form of
\begin{equation*}
\varphi^+ = c_1^+ e^{\Theta \xi_3} + c_2^+ e^{-\Theta \xi_3},\quad \varphi^- = c e^{\Theta \xi_3}.
\end{equation*}
The transmission conditions at $\xi_3=\eta_3$ give
\begin{equation}\label{is8}
\left\{
\begin{array}{l}
0=\varphi^+ - \varphi^-=c_1^+ e^{\Theta \eta_3} + c_2^+ e^{-\Theta \eta_3} - c e^{\Theta \eta_3}, \\
-e^{-(\tau s + i \zeta^\prime \cdot \eta^\prime)}=\Theta c_1^+ e^{\Theta \eta_3} - \Theta c_2^+ e^{-\Theta \eta_3} - \Theta c e^{\Theta \eta_3}.
\end{array}
\right.
\end{equation}
The boundary condition at $\xi_3=0$ leads to
\begin{equation*}
0=(\Theta c_1^+ - \Theta c_2^+) - \lambda_0 (c_1^+ + c_2^+)=(\Theta - \lambda_0)c_1^+ - (\Theta + \lambda_0)c_2^+,
\end{equation*}
which implies
\begin{equation}\label{is11}
c_2^+ = \frac{\Theta - \lambda_0}{\Theta + \lambda_0} c_1^+.
\end{equation}
From \eqref{is8} and \eqref{is11}, we obtain
\begin{equation*}
c_1^+ = \frac{1}{2} \Theta^{-1} e^{\Theta \eta_3} \frac{\Theta +\lambda_0}{\Theta -\lambda_0} e^{-(\tau s + i \zeta^\prime \cdot \eta^\prime)}, \quad
c_2^+ = \frac{1}{2} \Theta^{-1} e^{\Theta \eta_3} e^{-(\tau s + i \zeta^\prime \cdot \eta^\prime)}.
\end{equation*}
Thus, $\varphi^+$ is expressed by
\begin{equation}\label{is15}
\varphi^+ = \frac{1}{2\Theta} e^{-(\tau s + i\zeta^\prime\cdot \eta^\prime)} \left\{ e^{\Theta(\xi_3+\eta_3)} \frac{\Theta + \lambda_0}{\Theta - \lambda_0} + e^{\Theta (\eta_3-\xi_3)} \right\}.
\end{equation}
We note that in the brace of \eqref{is15} the second term comes from the fundamental solution, while the first term corresponds to the reflected solution $W^+$. So \eqref{is16} is obtained by taking the inverse Fourier and Laplace transforms of \eqref{is15}. The proof is complete. \hfill $\Box$

\bigskip
Finally, we show the pointwise asymptotic behavior of $W^+(\xi,\,t;\,\eta,\,s)$.

\begin{figure} [htp]
\begin{center}
\includegraphics[width=0.6\textwidth,height=6.0cm]{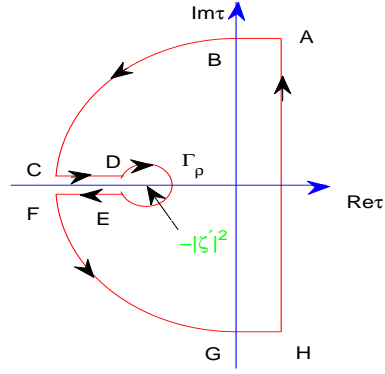}
\caption{In the contour marked with arrow, $R$ is the radius of the half circle and $\Gamma_{\rho}$ is the circle with radius $\rho$ centered at $(-|\zeta^\prime|^2,\,0)$.}\label{contour1}
\end{center}
\end{figure}

\begin{lemma}\label{thas2}
Let $\eta=(\eta_1,\,\eta_2,\,\eta_3)^T$, where $\eta_3=-\epsilon/2$ for small $\epsilon>0$. Then we have
\begin{equation}\label{is27}
W^+(\eta,\,s+\epsilon^2;\,\eta,\,s) = \frac{\epsilon^{-3}}{8e^{1/4}\pi^{3/2}} + O(\epsilon^{-2}) \thicksim \frac{\epsilon^{-3}}{8e^{1/4}\pi^{3/2}} \to \infty  \quad \textrm{ as } \epsilon\to 0.
\end{equation}
\end{lemma}
{\bf Proof.} To analyze the asymptotic behavior of $W^+$, let us first calculate
\begin{equation}\label{is17}
I=I(\zeta^\prime,\,t,\,s):=\int_{\sigma - i\infty}^{\sigma+i\infty} e^{(t-s)\tau} \frac{\Theta + \lambda_0}{\Theta(\Theta-\lambda_0)} e^{\Theta(\xi_3+\eta_3)} d\tau,\quad \xi_3,\,\eta_3<0
\end{equation}
by converting the line integral from $\sigma-i\infty$ to $\sigma+i\infty$ into a closed contour so that we can apply the residue theorem. The contour used here is described in Figure \ref{contour1}.

The contributions from the arcs $ABC$ and $FGH$ are negligibly small as $R\to \infty$ by Jordan's lemma. Let $\Gamma_\rho$ be the circle with radius $\rho$ centered at $(-|\zeta^\prime|^2,\,0)$. Then, by setting $\tau+|\zeta^\prime|^2=\rho e^{i\theta}$, we can show that
\begin{eqnarray}\label{is18}
&&\int_{\Gamma_\rho}e^{(t-s)\tau} \frac{\Theta + \lambda_0}{\Theta(\Theta-\lambda_0)} e^{\Theta(\xi_3+\eta_3)} d\tau\nonumber\\
&=& \int_{\pi}^{-\pi} e^{(t-s)(-|\zeta^\prime|^2+\rho e^{i\theta})} \frac{\sqrt{\rho}e^{i\theta/2}+\lambda_0}{\sqrt{\rho}e^{i\theta/2}(\sqrt{\rho}e^{i\theta/2}-\lambda_0)}e^{\Theta(\xi_3+\eta_3)}i\rho e^{i\theta} d\theta \to 0 \quad \textrm{as } \rho\to 0.
\end{eqnarray}
Let $R$ be the radius of the half circle in the contour. By taking $r=-|\zeta^\prime|^2-\tau$, we have
\begin{eqnarray}\label{is19}
&&\int_{\overrightarrow{CD}}e^{(t-s)\tau} \frac{\Theta + \lambda_0}{\Theta(\Theta-\lambda_0)} e^{\Theta(\xi_3+\eta_3)} d\tau\nonumber\\
&=&\int_{R-|\zeta^\prime|^2}^{0} e^{-(t-s)(r+|\zeta^\prime|^2)} \frac{i\sqrt{r}+\lambda_0}{i\sqrt{r}(i\sqrt{r}-\lambda_0)}e^{i\sqrt{r}(\xi_3+\eta_3)}(-dr)\nonumber\\
&=& i e^{-(t-s) |\zeta^\prime|^2} \int_0^{R-|\zeta^\prime|^2} e^{-(t-s)r}\frac{-r+2i\lambda_0 \sqrt{r}+\lambda_0^2}{\sqrt{r}(r+\lambda_0^2)} e^{i\sqrt{r}(\xi_3+\eta_3)}dr \nonumber\\
&\to& i e^{-(t-s) |\zeta^\prime|^2} \int_0^\infty e^{-(t-s)r}\frac{-r+2i\lambda_0 \sqrt{r}+\lambda_0^2}{\sqrt{r}(r+\lambda_0^2)} e^{i\sqrt{r}(\xi_3+\eta_3)}dr \quad \textrm{ as } R\to\infty.
\end{eqnarray}
Similar to the derivation of \eqref{is19}, we also have
\begin{eqnarray}\label{is20}
&&\int_{\overrightarrow{EF}}e^{(t-s)\tau} \frac{\Theta + \lambda_0}{\Theta(\Theta-\lambda_0)} e^{\Theta(\xi_3+\eta_3)} d\tau\nonumber\\
&\to& i e^{-(t-s) |\zeta^\prime|^2} \int_0^\infty e^{-(t-s)r}\frac{-r-2i\lambda_0 \sqrt{r}+\lambda_0^2}{\sqrt{r}(r+\lambda_0^2)} e^{-i\sqrt{r}(\xi_3+\eta_3)}dr \quad \textrm{ as } R\to\infty.
\end{eqnarray}
Notice that $(\lambda_0^2-|\zeta^\prime|^2,0)$ is contained in the interior of the contour. By the residue theorem, we get
\begin{eqnarray}\label{is21}
I&=&8\pi\lambda_0 ie^{(t-s)(\lambda_0^2- |\zeta^\prime|^2)} e^{\lambda_0(\xi_3+\eta_3)} \nonumber\\
&&-2ie^{-(t-s) |\zeta^\prime|^2}\mathrm{Re}\left( \int_0^\infty e^{-(t-s)r}\frac{-r+2i\lambda_0 \sqrt{r}+\lambda_0^2}{\sqrt{r}(r+\lambda_0^2)} e^{i\sqrt{r}(\xi_3+\eta_3)}dr\right).
\end{eqnarray}
Define
\begin{equation*}
L=L(t,\,s,\,\xi_3,\,\eta_3;\,\lambda_0):=\mathrm{Re}\left( \int_0^\infty e^{-(t-s)r}\frac{-r+2i\lambda_0 \sqrt{r}+\lambda_0^2}{\sqrt{r}(r+\lambda_0^2)} e^{i\sqrt{r}(\xi_3+\eta_3)}dr\right).
\end{equation*}
Then $W^+$ can be represented by
\begin{eqnarray}\label{is22}
W^+(\xi,\,t;\,\eta,\,s)&=&\frac{1}{16\pi^3i}\int_{\mathbb R^2}e^{i(\xi^\prime - \eta^\prime)\cdot \zeta^\prime} \, I d\zeta^\prime \nonumber\\
&=& \left(\frac{\lambda_0}{2\pi^2}e^{\lambda_0^2(t-s)+\lambda_0(\xi_3+\eta_3)}  -\frac{1}{8\pi^3} L \right)\int_{\mathbb R^2} e^{-(t-s)|\zeta^\prime|^2} e^{i(\xi^\prime - \eta^\prime)\cdot \zeta^\prime} d\zeta^\prime.
\end{eqnarray}

By direct calculations, we obtain that
\begin{eqnarray*}
\int_{\mathbb R} e^{-(t-s)\zeta_1^2} e^{i(\xi_1-\eta_1)\zeta_1} d\zeta_1
&=& (\sqrt{t-s})^{-1} \int_{\mathbb R} e^{-(\zeta_1^\prime)^2} e^{i(\sqrt{t-s})^{-1}(\xi_1 - \eta_1)\zeta_1^\prime} d\zeta_1^\prime\\
&=& (\sqrt{t-s})^{-1} e^{-\frac{1}{4}(t-s)^{-1}(\xi_1-\eta_1)^2} \int_{\mathbb R} e^{-(\tilde \zeta_1)^2} d \tilde \zeta_1\\
&=& \sqrt{\pi} (\sqrt{t-s})^{-1} e^{-\frac{1}{4}(t-s)^{-1}(\xi_1-\eta_1)^2}.
\end{eqnarray*}
Similarly, we have
\begin{equation*}
\int_{\mathbb R} e^{-(t-s)\zeta_2^2} e^{i(\xi_2-\eta_2)\zeta_2} d\zeta_2=\sqrt{\pi} (\sqrt{t-s})^{-1} e^{-\frac{1}{4}(t-s)^{-1}(\xi_2-\eta_2)^2}.
\end{equation*}
Therefore, it follows from \eqref{is22} that
\begin{equation}\label{is23}
W^+(\xi,\,t;\,\eta,\,s)= \left(\frac{\lambda_0}{2\pi (t-s)}e^{\lambda_0^2(t-s)+\lambda_0(\xi_3+\eta_3)} - \frac{1}{8\pi^2 (t-s)} L\right)e^{-\frac{1}{4}(t-s)^{-1}|\xi^\prime-\eta^\prime|^2}.
\end{equation}

\begin{figure} [htp]
\begin{center}
\includegraphics[width=0.6\textwidth,height=6.0cm]{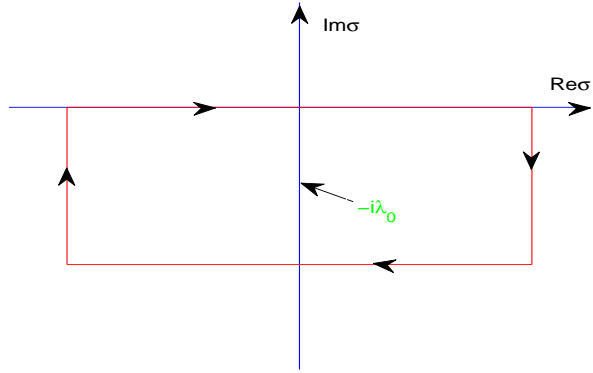}
\caption{The contour marked with arrow contains the point $(0,\,-\lambda_0)$ in its interior.}\label{contour2}
\end{center}
\end{figure}

In the sequel, we compute $L$. By setting $p=\sqrt{r}$, we have
\begin{eqnarray}\label{is24}
L&=& \mathrm{Re} \int_0^\infty e^{-(t-s)p^2} \frac{-p^2 + 2i\lambda_0 p + \lambda_0^2}{p (p^2+\lambda_0^2)} e^{ip (\xi_3 + \eta_3)} 2 p dp\nonumber\\
&=& \mathrm{Re} \int_{-\infty}^{+\infty} \frac{-p^2 + \lambda_0^2}{p^2 +\lambda_0^2} e^{-(t-s)p^2+ip(\xi_3+\eta_3)} dp - 2 \lambda_0\, \mathrm{Im}  \int_{-\infty}^{+\infty} \frac{p}{p^2+\lambda_0^2}  e^{-(t-s)p^2+ip(\xi_3+\eta_3)} dp.
\end{eqnarray}
To compute the integrals in \eqref{is24}, we introduce the contour described in Figure \ref{contour2} so that the residue theorem can be applied.

Note that
$$-(t-s)p^2 + ip(\xi_3+\eta_3)=-\left(\sqrt{t-s} \, p - \frac{1}{2}\frac{i}{\sqrt{t-s}}(\xi_3+\eta_3) \right)^2-\frac{(\xi_3+\eta_3)^2}{4(t-s)},$$
and define
$$\tilde p:=\sqrt{t-s} \, p - \frac{1}{2}\frac{i}{\sqrt{t-s}}(\xi_3+\eta_3).$$
It holds that
\begin{eqnarray*}
-p^2&=&-i(t-s)^{-1}(\xi_3+\eta_3)\left\{ (t-s)^{-1/2}\tilde p +\frac{i}{2}(t-s)^{-1}(\xi_3+\eta_3) \right\} - \frac{(\xi_3+\eta_3)^2}{4(t-s)^2}-{\tilde p}^2(t-s)^{-1}\\
&=&-i(t-s)^{-3/2}(\xi_3+\eta_3)\tilde p + \frac{1}{4}(\xi_3+\eta_3)^2(t-s)^{-2}-{\tilde p}^2(t-s)^{-1}\\
&=&A+ib\tilde p,
\end{eqnarray*}
where
$$A:=\frac{1}{4}(\xi_3+\eta_3)^2(t-s)^{-2} - \tilde p^2 (t-s)^{-1},\quad b:=-(t-s)^{-3/2}(\xi_3+\eta_3).$$

Using the residue theorem and noticing that the integrals on $\mathrm{Re}\,p=\pm R$ go to zero as $R\to \infty$, we obtain from \eqref{is24} that
\begin{eqnarray}\label{is25}
L&=&4\pi \lambda_0 e^{(t-s)\lambda_0^2} e^{\lambda_0(\xi_3+\eta_3)} +(t-s)^{-1/2} e^{-\frac{(\xi_3+\eta_3)^2}{4(t-s)}}\int_{-\infty}^{+\infty} e^{-{\tilde p}^2} \frac{-A^2+\lambda_0^4-b^2{\tilde p}^2}{(-A+\lambda_0^2)^2+b^2{\tilde p}^2}d\tilde p\nonumber\\
&&-\lambda_0 e^{-\frac{(\xi_3+\eta_3)^2}{4(t-s)}} \int_{-\infty}^{+\infty} e^{-{\tilde p}^2}  \frac{(A-\lambda_0^2)b+2b{\tilde p}^2(t-s)^{-1}}{(-A+\lambda_0^2)^2+b^2{\tilde p}^2} d\tilde p.
\end{eqnarray}
Now take $\xi=\eta$ with $\xi_3=\eta_3=-\epsilon/2$ and $t-s=\epsilon^2$. Then $b=\epsilon^{-2}$ and $A=\epsilon^{-2}(\frac{1}{4}-\tilde p^2)$. By direct calculations, we have
\begin{eqnarray*}
\frac{-A^2+\lambda_0^4-b^2{\tilde p}^2}{(-A+\lambda_0^2)^2+b^2{\tilde p}^2}&=& -1 + \epsilon^2\frac{2\lambda_0^2 \tilde p^2 +2 \epsilon^2 \lambda_0^4 -\lambda_0^2/2}{\tilde p^4 + (1/2 + 2 \epsilon^2 \lambda_0^2)\tilde p^2 + (\epsilon^2 \lambda_0^2 - 1/4)^2}, \\
\frac{(A-\lambda_0^2)b+2b{\tilde p}^2(t-s)^{-1}}{(-A+\lambda_0^2)^2+b^2{\tilde p}^2} & = & \frac{\tilde p^2 - \lambda_0^2 \epsilon^2 + 1/4}{\tilde p^4 + (1/2 + 2 \epsilon^2 \lambda_0^2)\tilde p^2 + (\epsilon^2 \lambda_0^2 - 1/4)^2}.
\end{eqnarray*}
Hence, it can be easily seen that
\begin{eqnarray*}
\int_{-\infty}^{+\infty} e^{-\tilde p^2} \frac{-A^2+\lambda_0^4-b^2{\tilde p}^2}{(-A+\lambda_0^2)^2+b^2{\tilde p}^2} d \tilde p &=& - \sqrt{\pi} + O(\epsilon^2), \\
\int_{-\infty}^{+\infty} e^{-\tilde p^2} \frac{(A-\lambda_0^2)b+2b{\tilde p}^2(t-s)^{-1}}{(-A+\lambda_0^2)^2+b^2{\tilde p}^2} d \tilde p  & = & O(1)
\end{eqnarray*}
for $\epsilon \to 0$. Consequently, we have
\begin{equation*}
L= 4\pi \lambda_0 - e^{-1/4}\sqrt{\pi} \epsilon^{-1} + O(1) \quad \textrm{as }\,\epsilon \to 0.
\end{equation*}
Thus, in terms of \eqref{is23}, we finally obtain
\begin{equation*}
W^+(\eta,\,s+\epsilon^2;\,\eta,\,s) = \frac{\epsilon^{-3}}{8e^{1/4}\pi^{3/2}} + O(\epsilon^{-2}) \thicksim \frac{\epsilon^{-3}}{8e^{1/4}\pi^{3/2}} \to \infty  \quad \textrm{ as } \epsilon\to 0.
\end{equation*}
This completes the proof. \hfill $\Box$

\medskip
Based on the above lemmas, we conclude that $W^+$ is the dominant part of $\tilde R$ as $\epsilon\to 0$. More explicitly, the pointwise asymptotic behavior of $\tilde R(\xi,\,t;\,\eta,\,s)$ can be stated as follows.
\begin{theorem}
Let $\eta=(\eta_1,\,\eta_2,\,\eta_3)^T$, where $\eta_3=-\epsilon/2$ for small $\epsilon>0$. Then we have
\begin{equation}\label{final}
\tilde R(\eta,\,s+\epsilon^2;\,\eta,\,s) = \frac{\epsilon^{-3}}{8e^{1/4}\pi^{3/2}} + O(\epsilon^{-2}) \thicksim \frac{\epsilon^{-3}}{8e^{1/4}\pi^{3/2}} \to \infty  \quad \textrm{ as } \epsilon\to 0.
\end{equation}
\end{theorem}

\begin{remark}
For any fixed discrepancy in \eqref{eq:data1}, combining \eqref{asi12} with \eqref{final}, we finally have the following asymptotic behavior
\begin{equation}
(Sg^y)(y,\,s+\epsilon^2)=\frac{\epsilon^{-3}}{8e^{1/4}\pi^{3/2}} + O(\epsilon^{-2}) \quad \textrm{ as } \epsilon\to 0.
\end{equation}

\end{remark}

\section{Concluding remarks}\label{con}

This paper investigated the inverse problem of identifying an unknown Robin-type cavity inside a heat conductor from boundary measurements. The so-called linear sampling method was established to reconstruct the shape and location of the cavity. Based on our well-posedness analysis of the corresponding forward problem, we gave rigorously a theoretical justification of this reconstruction scheme by using the layer potential argument. Further, we proved a short time asymptotic behavior of the reflected solution of the fundamental solution, and hence the asymptotic behavior of the indicator function used in this method was obtained. Since the reflected solution is the compensating term of the Green function for the related initial-boundary value problem, we actually provided a short time asymptotic behavior of the Green function. The asymptotic behavior naturally yields a pointwise reconstruction scheme for the boundary of the cavity. We would also like to emphasize that from the asymptotic behavior we can know the distance to the unknown cavity as we probe it from its inside. However, to establish a pointwise reconstruction formula for the Robin coefficient $\lambda$, we need to carefully examine the lower order term where the information about the Robin coefficient should be involved. This is one of our future works. Also, we intend to utilize this asymptotic behavior to generate a good numerical performance of this reconstruction scheme.

\bigskip
{\bf Acknowledgement:} The first author is partially supported by National Research Foundation of Korea (No. 49771-01). The second author is supported by National Natural Science Foundation of China (Nos. 11301075, 91330109) and Natural Science Foundation of Jiangsu Province of China (No. BK20130594).

\end{document}